\documentclass[aps,pra,twocolumn,showpacs,superscriptaddress,longbibliography]{revtex4-2}
\usepackage{graphicx}    
\usepackage{dcolumn}    
\usepackage{bm}             
\usepackage{amssymb}   
\usepackage{amsmath}    
\usepackage{subfigure}    
\usepackage{bbm}
\usepackage{bbding} 
\usepackage{physics}
\usepackage{braket}
\usepackage{mathrsfs}
\usepackage{multirow}
\usepackage[export]{adjustbox}
\usepackage[toc,page]{appendix}
\usepackage{booktabs} 
\begin{document}

\title{Subspace Leakage Error Randomized Benchmarking of M{\o}lmer-S{\o}rensen Gates}
\author{R.~T.~Sutherland}
\affiliation{Oxford Ionics (an IonQ company), Oxford, OX5 1PF, UK}
\author{A.~C.~Hughes}
\affiliation{Oxford Ionics (an IonQ company), Oxford, OX5 1PF, UK}
\author{J.~P.~Marceaux}
\affiliation{Oxford Ionics (an IonQ company), Oxford, OX5 1PF, UK}
\author{H.~M.~Knaack}
\affiliation{Oxford Ionics (an IonQ company), Oxford, OX5 1PF, UK}
\author{C.~M.~L{\"o}schnauer}
\affiliation{Oxford Ionics (an IonQ company), Oxford, OX5 1PF, UK}
\author{R.~Srinivas}
\email{r.srinivas@oxionics.com}
\affiliation{Oxford Ionics (an IonQ company), Oxford, OX5 1PF, UK}
\affiliation{Department of Physics, University of Oxford, Oxford, OX1 3PU, UK}

\date{\today}

\begin{abstract}
We demonstrate a new technique that adapts single-qubit randomized benchmarking to two-qubit M{\o}lmer-S{\o}rensen gates. We use the controllable gate phase to generate Cliffords that act on a two-state subspace, enabling benchmarking of two-qubit gates without single-qubit operations. In addition to quantifying the gate infidelity, the protocol provides valuable information about the type of error by distinguishing between those that conserve the two-state subspace and those that result in leakage out of it. We demonstrate the protocol for calibrating and validating all-electronic maximally entangling gates in a trapped-ion quantum computer, achieving a two-qubit gate error of $2.6 (2)\times10^{-4}$.
\end{abstract}
\maketitle

\section{Introduction}

One of the most important metrics describing a quantum computer is the fidelity of two-qubit entangling operations~\cite{preskill_2018}. While two-qubit gates can already be performed below the quantum error correction (QEC) threshold in many physical platforms~\cite{shor_1995, steane_1996, calderbank_1996, gottesman_1997}, further gate fidelity improvements are still highly beneficial, as they allow for reduced QEC overheads, additional near-term applications, and increased engineering margin during QC scale-up. To that end, we need methods that estimate gate errors \textit{and} methods that characterize the noise mechanisms underlying the errors. Such methods also need to scale easily to lower error rates as two-qubit gate fidelities continue to improve. 

The popular approach to gate quality characterization is through randomized benchmarks  -- a family that includes, among others, Clifford randomized benchmarking (RB)~\cite{knill_2008, ryan_2009, gaebler_2012}, character RB~\cite{helsen_new_2019}, cross-entropy benchmarking~\cite{arute_2019}, mirror circuit benchmarks~\cite{Proctor_2022}, and quantum volume~\cite{Cross_2019}.  Clifford RB is particularly efficient as it comprises a series of random Clifford operators that form a two-design on the subspace they act on, and can be simulated classically. Clifford RB can be used to measure arbitrarily small errors by increasing the sequence length, and measurement at different sequence lengths can also eliminate state preparation and measurement (SPAM) errors. However, as Clifford RB depolarises all error channels by design, it provides no information about the error sources. Further, the required Clifford operations are composed of both two-qubit and addressed single-qubit gates, making it challenging to disentangle errors associated with two-qubit rotations, single-qubit rotations, and individual addressing. Symmetric-subspace benchmarking~\cite{baldwin_2020} eliminates the need for individual addressing, but still requires single-qubit rotations. This requirement poses a limitation, especially as two-qubit gate error rates continue to improve~\cite{loeschnauer_2024}.

On the other hand, many techniques have been developed to characterize the nature of errors in two-qubit gates. Historically, partial state tomography (PST) was used in the early entanglement demonstrations~\cite{sackett_2000}. While PST is sensitive to SPAM errors, and -- as it only contains a single entangling gate -- has high statistical uncertainty per shot, separate measurements of the populations and coherences help distinguish between different error mechanisms. In general, tomographic approaches such as direct fidelity estimation~\cite{Flammia_2011}, process tomography~\cite{Chuang_1997}, and gate set tomography~\cite{Nielsen_2021} can offer even greater insight into error processes. However, they require a larger number of measurements and have a more complex data analysis procedure. It would be useful to combine the scalability, SPAM insensitivity, and simplicity of Clifford RB, with the ability to pinpoint the origins of error offered by tomographic techniques. \\

In this paper, we propose and demonstrate one such method. Our technique is specifically aimed at characterizing two-qubit M{\o}lmer-S{\o}rensen gates~\cite{molmer_1999, molmer_2000} in trapped-ion systems, and it uses the tunable MS gate phase degree of freedom $\phi_{\text{MS}}$ as the singular free parameter in each circuit decomposition. Varying $\phi_{\text{MS}}$ allows us to adapt single-qubit RB to generate Cliffords on a two-state subspace $\mathcal{S}_{\text{RB}}$ of the full two-qubit subspace $\mathcal{S}_{2Q}$. Measuring the rate of population transfer, both within $\mathcal{S}_{\text{RB}}$ and between $\mathcal{S}_{\text{RB}}$ and other subspaces $\mathcal{S}_{\text{leak}}\subset\mathcal{S}_{2Q}$, gives us an estimation of the average two-qubit gate fidelity and provides useful information for characterizing errors. Hence, we call the technique subspace leakage error randomized benchmarking (SLERB). 

This paper is structured as follows. Section~\ref{sec:theory} describes the theory of SLERB using two approaches. One approach describes the technique by examining the relationship between certain population transfer matrix elements and the average gate infidelity (Sec.~\ref{sec:transfer_matrix}), and the other examines the technique's group-theoretical properties (Sec.~\ref{sec:group_theory}). In Section~\ref{sec:sims}, we perform numerical simulations of SLERB sequences to verify that it faithfully reproduces true gate errors. Finally, in Sec.~\ref{sec:implementation}, we demonstrate SLERB for calibrating and characterizing all-electronic MS gates on trapped-ion qubits, measuring an error rate of $2.6 (2)\times10^{-4}$.
 
\begin{figure}[!b]
\includegraphics[width=1\columnwidth]{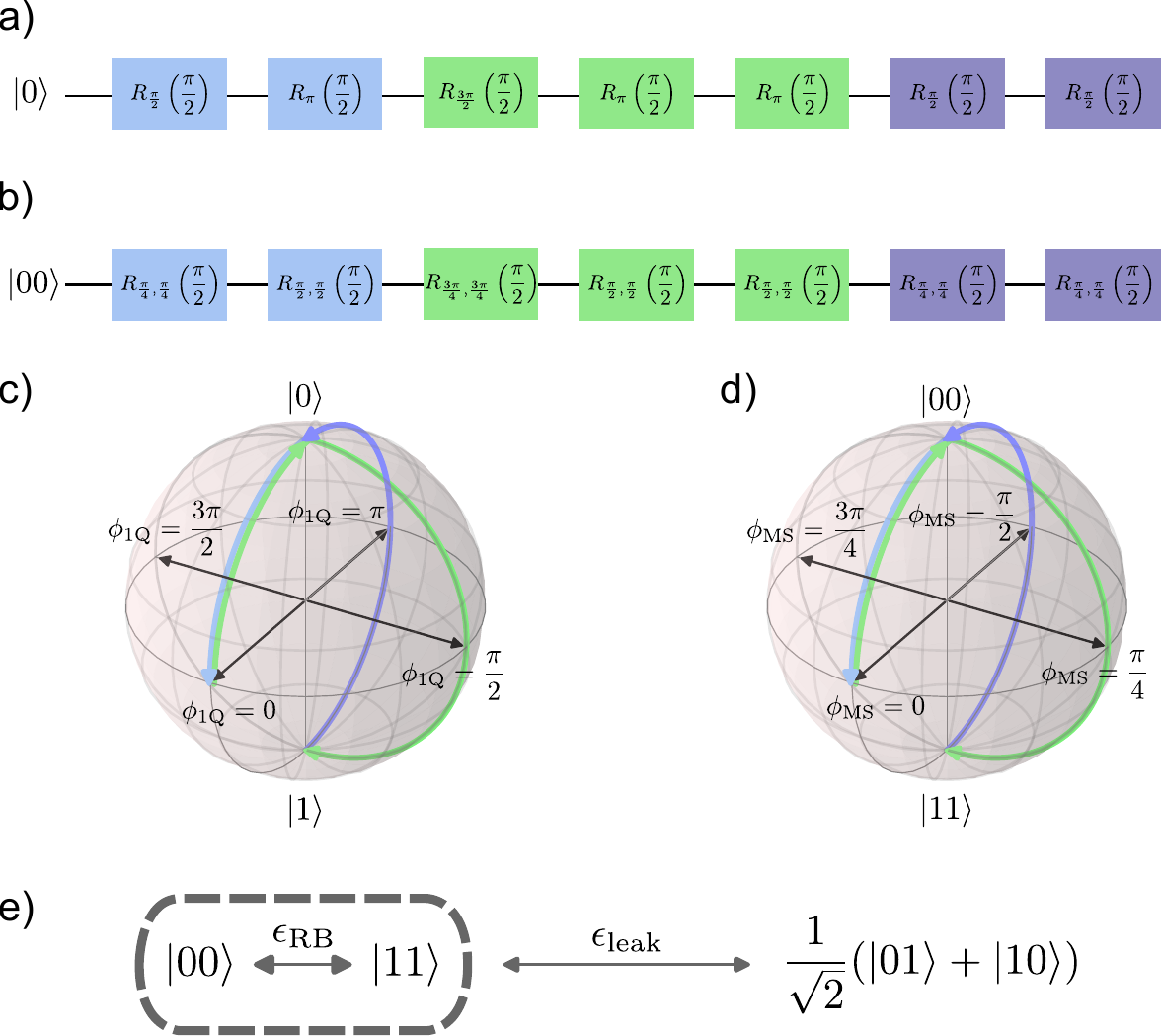}
\centering
\caption{\label{fig_schematic} Conversion of single-qubit to two-qubit sequences. SLERB sequences are generated from (a) sequences of single-qubit rotations ($R_\phi(\pi/2)$, $\phi=\phi_\textrm{1Q}$) (a), which are converted to (b) sequences of two-qubit MS gates, ($R_{\phi,\phi}(\pi/2)$, $\phi=\phi_\textrm{MS}=\phi_\textrm{1Q}/2$). Each pulse ideally rotates the states by an angle $\pi/2$ within their respective two-state subspaces, as shown in the Bloch spheres for (c) single and (d) two qubits. Each rotation is color-matched to its respective Clifford in (a) and (b), and corresponds to a specific axis of the Bloch sphere specified by $\phi_\textrm{1Q}$ or $\phi_\textrm{MS}$. (e) Overview of population transfers. SU2 errors keep the states within the $\{\ket{00},\ket{11}\}$ subspace, with a corresponding error rate $\epsilon_\textrm{RB}$, while leakage errors result in populations in the $\frac{1}{\sqrt{2}}(\ket{01}+\ket{10})$ state with a corresponding error rate $\epsilon_\textrm{leak}$.}
\end{figure}

\section{Theory}\label{sec:theory}

\subsection{General overview}

\subsubsection{Fidelity estimation techniques }
In the context of characterizing single- and two-qubit gates, Clifford RB has become a ubiquitous, industry-standard technique. Clifford RB ``twirls'' coherent gate noise into uniform depolarization noise that can be simply characterized. The fundamental mathematical reasoning is that sampling from a unitary two-design allows one to approximate the Haar measure~\cite{Dankert_2009}. The experimental implementation involves constructing physical gate sequences that ideally compile to representations of elements of the Clifford group on $n$-qubits. A circuit is constructed by selecting a random number of $l$ Cliffords plus an $(l+1)$th inversion operation, and the whole sequence is then run after preparing a given initial state. The resulting success probability $P_\text{survival}$ of measuring the input state at the output can be shown to decay according to a single exponential in the length $l$ of the Clifford sequence
\begin{equation}
    P_\text{survival} = A + B q^{l}_{\text{RB}}. 
\end{equation}
Here, $q_{\text{RB}}$ is an experimentally observed decay parameter, and $A, B \leq 1$ are constants that encode the effects of state preparation and measurement (SPAM) errors as well as any errors in the final inverting Clifford. Under certain assumptions, there is a strong relationship between $q_{\text{RB}}$ and the fidelity of the gates used to generate the Clifford sequence~\cite{Helsen_2022}. 

Clifford RB has been generalized to address non-standard noise effects such as leakage. Importantly for our work, Ref.~\cite{Wood_2018} generalized single-qubit Clifford RB to include the effects of leakage from the qubit manifold to neighboring states. In the case of only one leakage process, the resulting success probability is expected to decay according to a double exponential model 
\begin{equation}
   P_\text{survival} = A + B q^{l}_{\text{RB}} + Cq_{\text{leak}}^l. 
\end{equation}
The decay in the case of leakage to many states can have more exponential terms -- one for each type of leakage process. Similar work~\cite{Andrews_2019} introduced a ``Blind RB'' protocol that isolates single-exponential decays by taking linear combinations of $P_\text{survival}$ and the bit flip probability $P_\text{flip}$, as 
\begin{align}
\label{eq:blind_rb}
    P_\text{survival} + P_\text{flip} &= 2(A + Cq_{\text{leak}}^l), \nonumber \\
    P_\text{survival} - P_\text{flip} &= 2 B q_{\text{RB}}^l.
\end{align}
The technique we introduce is analogous to these protocols applied to a single-qubit subspace of a two-qubit gate. 

Standard Clifford RB has also been modified to address particular experimental or platform-specific desiderata. Especially relevant for this work, Ref.~\cite{baldwin_2020} derived a protocol that implements an effective qutrit Clifford RB protocol~\cite{Morvan_2021} on the symmetric subspace (here, symmetric refers to qubit exchange) of a two-qubit Hilbert space in trapped ions. This protocol removes a number of experimental confounders, notably that antisymmetric operations in some trapped ion systems require transport, which might introduce new sources of error that are independent of the performance of the two-qubit gate.

\subsubsection{Subspace leakage error randomized benchmarking}\label{sec:slerb_theory_description}

Figure~\ref{fig_schematic} shows how our protocol builds upon single-qubit randomized benchmarking (1QRB), where a sequence of operations sampled at random from the set of single-qubit Cliffords $\mathcal{C}_{1Q}$ is used to estimate the error of single-qubit gates. Each $\hat{C}_{1Q}\in\mathcal{C}_{1Q}$ can be decomposed into a series of single-qubit gates driving $R_{\phi_\textrm{1Q}}(\pi/2)$ rotations around the $x$, $y$, $-x$, and $-y$ axes on the Bloch sphere~\cite{feynman_1957}, with the rotation axis defined by $\phi_\textrm{1Q}$. This sequence of single-qubit rotations is converted to two-qubit gates, driving $R_{\phi_\textrm{MS, 1},\phi_\textrm{MS, 2}}(\pi/2)$ rotations, except the rotation axis is set by $\phi_\textrm{MS}=\phi_\textrm{1Q}/2$. Figure~\ref{fig_schematic}(d) shows how these pulses can be visualized on the $\mathcal{S}_{\text{RB}}\equiv \{\ket{00},\ket{11}\}$ Bloch sphere.

We implement our benchmarking protocol following the approach of single-qubit RB~\cite{knill_2008}. Explicitly, we identify 24 MS pulse sequences that compile to $\mathcal{C}_{1Q}$ when acting on $\mathcal{S}_{\text{RB}}$. We select a sequence of $l$ Cliffords at random from the set of 24 Cliffords and then calculate the inverse of the sequence. We then implement the physical pulse sequence for the $l+1$ Clifford gates, including the final inverse operation, on an initial state of $\ket{00}$. In this way, we produce a final state $\rho^{(l)}$ according to the map 
\begin{equation}
    \rho^{(l)} = G_{l+1} \circ G_{l} \circ ... \circ G_{1}\left[ |00\rangle \langle 00| \right],
\end{equation}
where $G_k$ is a channel representation of the $k$th noisy Clifford. Finally, we measure $\rho^{(l)}$ in the computational basis. We repeat the procedure to collect a number of shots $s$ from a particular compilation, and we further run the protocol for a number $r$ of different randomizations in the sequences. In this way, we collect $s \times r$ random categorical variables supported on the measurement outcome space $\{00, 01, 10, 11\}$. This protocol is a direct generalization of single-qubit RB. From this perspective, we can directly deploy standard approaches to calculate an effective SU(2) error rate on the embedded subspace of the MS gate used in SLERB. However, it is desirable to extend our results to estimate gate errors averaged over $\mathcal{S}_{2Q}$. 

\subsubsection{Clifford decompositions}\label{sec:clifford_decompositions}

A `standard' two-qubit Clifford RB circuit comprises two-qubit Clifford operators $\mathcal{C}_{2Q}$. However, as generating $\mathcal{C}_{2Q}$ on a trapped-ion QC involves many physical operations in addition to two-qubit gates (addressed single-qubit gates, transport, cooling etc.), this creates challenges for directly estimating the two-qubit gate error rate. Ref.~\cite{baldwin_2020} partially addressed this issue with an RB variant they referred to as `subspace randomized benchmarking' (SRB). The technique uses benchmarking sequences comprising symmetric (with respect to qubit exchange) $d=3$ Cliffords $\mathcal{C}_{+}$. However, SRB still requires single-qubit gates in addition to two-qubit gates.

With this in mind, we focus on sequences that decompose into only the unitaries naturally generated by the M\o lmer-S\o rensen (MS) gate~\cite{molmer_1999,molmer_2000}. Up to a global phase, this can be expressed as:
\begin{eqnarray}\label{eq:ideal_unitary}
    \hat{U}_{2Q}(\theta_{\text{MS}},\phi_{\text{MS}}) &=& \exp\Big(-\frac{i\theta_{\text{MS}}}{2}\hat{\sigma}_{\phi_{\text{MS}},1}\hat{\sigma}_{\phi_{\text{MS}},2} \Big),
\end{eqnarray}
where $\hat{\sigma}_{\phi_{\text{MS}}, j}\equiv \hat{\sigma}_{x, j}\cos(\phi_{\text{MS}})+\hat{\sigma}_{y,j}\sin(\phi_{\text{MS}})$ is a Pauli operator acting on qubit $j$. We choose $\phi_{\text{MS}}$ to be the free parameter to generate our Clifford decompositions; this choice is convenient for MS gate experiments because $\phi_{\text{MS}}$ is the phase of the gate fields relative to the qubit and is relatively straightforward to adjust.  However, this method can also be applied to $\propto \hat{\sigma}_{z,1}\hat{\sigma}_{z,2}$ gates by adding collective single-qubit $\pi/2$-rotations before and after the gate.

The `even' $\mathcal{S}_{\text{RB}}\equiv \{\ket{00},\ket{11} \}$ and `odd' $\mathcal{S}_{\text{leak}}\equiv \{\ket{01},\ket{10} \}$ subspaces are invariant under $\hat{U}_{2Q}$; note that $\mathcal{S}_{\text{RB}}+\mathcal{S}_{\text{leak}}=\mathcal{S}_{2Q}$, where $\mathcal{S}_{2Q}$ is the total two-qubit Hilbert space. In the following, it will be convenient to further divide $\mathcal{S}_{\text{leak}}$ into symmetric $\mathcal{S}_{\text{leak},+}\equiv \{2^{-1/2}(\ket{01}+\ket{10}) \}$ and antisymmetric $\mathcal{S}_{\text{leak}^{-}}\equiv \{2^{-1/2}(\ket{01}-\ket{10}) \}$ single-state subspaces. The action of $\hat{\sigma}_{\phi_{\text{MS}},1}\hat{\sigma}_{\phi_{\text{MS}},2}$ on $\mathcal{S}_{\text{RB}}$ is:
\begin{eqnarray}\label{eq:isometry_example}
    \hat{\sigma}_{\phi_{\text{MS}},1}\hat{\sigma}_{\phi_{\text{MS}},2}\Big |_{S_{\text{RB}}} &=& e^{-2i\phi_{\text{MS}}}\ket{11}\bra{00}+e^{2i\phi_{\text{MS}}}\ket{00}\bra{11}, \nonumber \\
    &\equiv& \hat{\sigma}_{2\phi_{\text{MS}},\text{RB}}.
\end{eqnarray}
Thus, $\hat{U}_{2Q}$ is isomorphic to a single-qubit Pauli $\hat{\sigma}_{2\phi_{\text{MS}},\text{RB}}$ operator when acting on $\mathcal{S}_{\text{RB}}$. Similarly, the collective Pauli-$z$ operator:
\begin{eqnarray}
\frac{1}{2}(\hat{\sigma}_{z,1}+\hat{\sigma}_{z,2})\Big|_{\mathcal{S}_{\text{RB}}}\equiv \hat{\sigma}_{z,\text{RB}},    
\end{eqnarray}
is isomorphic to a single-qubit Pauli-$z$ operator when acting on $\mathcal{S}_{\text{RB}}$. Since we can make effective $\propto \hat{\sigma}_{x,\text{MS}}$ and $\propto \hat{\sigma}_{y,\text{MS}}$ rotations on $\mathcal{S}_{\text{RB}}$, we can generate Euler rotations, and, therefore, any SU(2) operation on $S_{\text{RB}}$. Hence, we can generate any $\hat{C}_{1Q}$ on $\mathcal{S}_{\text{RB}}$, and implement single-qubit RB on $\mathcal{S}_{\text{RB}}$ with only $\hat{U}_{2Q}$ operations.

\subsection{Transfer matrix}\label{sec:transfer_matrix}

In a benchmarking sequence, an `ideal' circuit comprising $l$ random Cliffords will map the qubit onto some target `survival' state $\ket{\text{survival}_{l}}\in\mathcal{S}_{\text{RB}}$, i.e., the state generated if no errors occurred. In this section, we are interested in the population transfer matrix that results specifically from \textit{errors} in each Clifford in this series. In the experiment, we measure the average population vector after $l$ Cliffords $\vec{P}_{l}\equiv (P_{\text{survival},l},P_{\text{flip},l},P_{\text{leak},l})$. Here, $P_{\text{survival},l}$ is the probability of measuring $\ket{\text{survival}_{l}}$ after the $l$ Clifford sequence, $P_{\text{flip},l}$ is the population of $\mathcal{S}_{\text{RB}}-\{\ket{\text{survival}_{l}}\}$ the other state in the RB subspace, and $P_{\text{leak},l}$ is the population in $\mathcal{S}_{\text{leak}}$ the total leakage subspace. We define the transfer matrix $T$ with the equation:
\begin{eqnarray}
    \vec{P}_{l+1}&=&(I+T)\vec{P}_{l},
\end{eqnarray} 
where $T=0$ indicates no population transfer errors. We want to determine the matrix elements of $T$. Let $k\in\mathcal{K}$ be the set of all noise sources $k$ present during the gate, all of which are assumed to be small. We can calculate the population transfer resulting from each noise mechanism $k$ (in isolation), then add them to obtain the transfer matrix resulting from all of $\mathcal{K}$:
\begin{eqnarray}
    T=\sum_{k\in\mathcal{K}}T_{k}.
\end{eqnarray}
For a given $k$ the total Hamiltonian is:
\begin{eqnarray}
    \hat{H}_{t}&=& \hat{H}_{g}+\hat{H}_{e,k},
\end{eqnarray}
where $\hat{H}_{g}$ is the ideal gate Hamiltonian and $\hat{H}_{e,k}$ is the Hamiltonian describing error source $k$. In the limit that $\hat{H}_{e,k}$ is small we can factor the total propagator as~\cite{blume_2022}: 
\begin{eqnarray}\label{eq:total_unitary_factored_error}
    \hat{U}_{t} &= & \hat{U}_{g}\tilde{U}_{e,k},
\end{eqnarray}
where $\tilde{U}_{e,k}$ is the time-propagator of $\tilde{H}_{e,k}$, which can be calculated perturbatively~\cite{sutherland_2022_1}. Since it is unitary, we can express the error unitary as:
\begin{eqnarray}
    \tilde{U}_{e,k} &\simeq & e^{-i\alpha_{k}\hat{Q}_{k}\otimes\hat{M}_{k}},
\end{eqnarray}
for some real number $\alpha_{k}$ and some Hermitian operator $\hat{Q}_{k}\otimes\hat{M}_{k}$, which we assume can be written as a tensor product of $\hat{Q}_{k}$ acting on the qubits and $\hat{M}_{k}$ acting on $\ket{m}$.

After $l$ Cliffords, the system will be in some mixed state:
\begin{eqnarray}\label{eq:rho_n}
\rho^{(l)}&=&\sum_{j}P_{j,l}P_{m,l}\ket{\psi_{j,l}}\bra{\psi_{j,l}}\otimes\ket{m}\bra{m},
\end{eqnarray}
which we interpret (equivalently) as the system having probability $P_{j,l}P_{m}$ of being in the pure state $\ket{\psi_{j,l}}\ket{m}$, where $\ket{\psi_{j,l}}$ refers to the qubits and $\ket{m}$ is an index that runs over the (potentially multiple) external degrees-of-freedom in the system. We assume the qubits are not entangled to any external degree of freedom after $l$ Cliffords. Eq.~\ref{eq:rho_n} depends both on the distribution (set of incoming states and their probabilities) of qubit states $\mathcal{D}_{l}\equiv \{\ket{\psi_{j,l}},P_{j,l} \}$ and external degrees-of-freedom $\mathcal{M}\equiv \{\ket{m},P_{m}\}$ that are being averaged over; all of the information in $\rho^{(l)}$ about the qubit states is contained in $\mathcal{D}$ and all the information about the external degrees-of-freedom are in $\mathcal{M}$. The reason we have included the index $l$ in $\ket{\psi_{j,l}}$ is because target $\ket{\text{survival}_{l}}$ and $\ket{\text{flip}_{l}}$ can change after each gate. The \textit{average} probability $P_{g,k}$ that the next $\hat{U}_{g}$ generates `the target state $\ket{g_{j,l+1}}$, conditioned on input state $\ket{\psi_{j,l}}$', i.e. $\ket{g_{j,l+1}}=\hat{U}_{g}\ket{\psi_{j,l}}$ even when $\ket{\psi_{j,l}}\neq\ket{\text{survival}_{l}}$, is:
\begin{eqnarray}\label{eq:prob_not_transfer}
    P_{g,k} &\equiv &\sum_{m^{\prime},m}\sum_{j}P_{j}P_{m}\Big|\bra{\psi_{j,l}}\bra{m^{\prime}}e^{-i\alpha_{k}\hat{Q}_{k}\otimes\hat{M}_{k}}\ket{\psi_{j,l}}\ket{m} \Big|^{2} \nonumber \\
    &\simeq & 1- \lambda^{2}_{\hat{Q}_{k}}\gamma_{k},
\end{eqnarray}
where:
\begin{eqnarray}
    \gamma_{k}\equiv |\alpha_{k}|^{2}\sum_{m}P_{m}\braket{m |\hat{M}_{k}^{2}|m},
\end{eqnarray}
quantifies the size of the noise channel, and
\begin{eqnarray}
    \lambda_{\hat{Q_{k}}}^{2}\!\!\!\equiv \!\!\sum_{j}\!P_{j}\Big(\!\!\braket{\psi_{j,l}|\hat{Q}^{2}_{k}|\psi_{j,l}}\!-\!\Big|\!\braket{\psi_{j,l}|\hat{Q}_{k}|\psi_{j,l}}\!\Big|^{2} \Big),
\end{eqnarray}
is the variance of $\hat{Q}_{k}$, averaged over $\mathcal{D}$. Eq.~\ref{eq:prob_transfer} is important because it shows we can factor out $E_{g,k}$'s dependence on $\mathcal{D}$. While this tells us the probability $E_{g,k}\equiv 1-P_{g,k}$ that a gate error results in population transfer to some state other than $\ket{g_{j,l}}$:
\begin{eqnarray}\label{eq:prob_transfer}
    E_{g,k} &\simeq & \lambda^{2}_{\hat{Q}_{k}}\gamma_{k},
\end{eqnarray}
we still need to know where this errant population will go before we can write an expression for $T_{k}$. 

We divide the set of all noise sources $\mathcal{K}=\mathcal{K}_{\text{RB}}+\mathcal{K}_{\text{leak}}$ into two subsets: noise that conserves $\mathcal{S}_{\text{RB}}$, which we label $\mathcal{K}_{\text{RB}}$, and noise that does not, which we will label $\mathcal{K}_{\text{leak}}$. The classification of each $k$ set by the structure of $\hat{Q}_{k}$; if needed for clarity, we the indicate the category of $\hat{Q}_{k}$ with subscripts $\hat{Q}_{\text{RB},k}$ or $\hat{Q}_{\text{leak},k}$. It is possible that some noise mechanism results in a $\hat{Q}_{k}$ with components that fall into both categories. In this case, we can treat $\hat{Q}_{k}$ as two distinct noise mechanisms $\hat{Q}_{\text{RB},k}$ and $\hat{Q}_{\text{leak},k}$. We assume our error budget comprises only symmetric errors, see Sec.~\ref{sec:symmetry}. If $\ket{\psi_{j,l}}\in\mathcal{S}_{\text{RB}}$, errors from $\mathcal{K}_{\text{RB}}$ will flip the state of $\ket{g_{j,l+1}}$ to the remaining state in $\mathcal{S}_{\text{RB}}-\ket{g_{j,l+1}}$. If $\ket{\psi_{j,l}}\in\mathcal{S}_{\text{leak}}$, then we assume $\mathcal{K}_{\text{RB}}$ does not affect $\ket{g_{j,l+1}}$. We can express every qubit operator resulting from $\mathcal{K}_{\text{RB}}$ as:
\begin{eqnarray}
    \hat{Q}_{k}\Big|_{\mathcal{S}_{\text{RB}}}&=& c_{i,k}\hat{I}_{\text{RB}} + c_{p,k}\vec{\sigma}_{\text{RB}}\cdot\hat{r},
\end{eqnarray}
noting we can incorporate any normalization into $|\alpha_{k}|^{2}$ that is necessary to ensure $c_{i,k}^{2}+c_{p,k}^{2}=1$. Since our gate set forms a two-design on $\mathcal{S}_{\text{RB}}$, we know the variances will be identical to the single-qubit Pauli operators averaged over SU(2):
\begin{eqnarray}
    \braket{\hat{\lambda}^{2}_{\hat{Q}_{k}}}_{\mathcal{S}_{\text{RB}}}&=& c^{2}_{p}\braket{\lambda^{2}_{\hat{Q}_{k}}}_{\mathcal{S}_{\text{RB}}} \nonumber \\
    &=& \frac{2}{3}c_{p}^{2}.
\end{eqnarray}
The only symmetric, $\mathcal{S}_{\text{RB}}$ conserving $\hat{Q}_{\text{RB}}$ that is $\propto \hat{I}_{\text{RB}}$ on $\mathcal{S}_{\text{RB}}$ is $\hat{\sigma}_{z,1}\hat{\sigma}_{z,2}$, while, for every other $\hat{Q}_{\text{RB}}\in\mathcal{Q}_{\text{RB}}$, $c_{p}^{2}=1$. Thus, $\propto \hat{\sigma}_{z,1}\hat{\sigma}_{z,2}$ error channels will be invisible to SLERB, and do not enter into our infidelity estimate; for the $\propto \hat{\sigma}_{\phi,1}\hat{\sigma}_{\phi,2}$ MS gates, error channels of this form are likely to be small. For every other symmetric $\hat{Q}_{\text{RB}}$, $c_{p}^{2}=1$, which we will assume from here on. We can now write an expression for the total rate of population transfer between the states in $\mathcal{S}_{\text{RB}}$:
\begin{eqnarray}
    \epsilon_{\text{RB}}&=& \frac{2}{3}\sum_{k\in\mathcal{K}_{\text{RB}}}\gamma_{k},
\end{eqnarray}
giving the $\mathcal{S}_{\text{RB}}$ conserving component of the transfer matrix:
\begin{equation}
    T_{\text{RB}} = \begin{bmatrix}
        -\epsilon_{\text{RB}} & \epsilon_{\text{RB}} & 0  \\
        \epsilon_{\text{RB}} & -\epsilon_{\text{RB}} & 0   \\
        0 & 0  & 0  \\
    \end{bmatrix}.
\end{equation}
Next, consider the set of two-qubit Pauli operators that couple $\mathcal{S}_{\text{RB}}$ and $\mathcal{S}_{\text{leak}}$:
$\mathcal{Q}_{\text{leak}}\equiv \{(\hat{\sigma}_{\phi,1}+\hat{\sigma}_{\phi,2}),( \hat{\sigma}_{\phi,1}\hat{\sigma}_{z,2}+\hat{\sigma}_{z,1}\hat{\sigma}_{\phi,2})\}$. Each operator in $\mathcal{Q}_{\text{leak}}$ couples both the states in $\mathcal{S}_{\text{RB}}$ to the single-state subspace $\mathcal{S}_{\text{leak},+}$. The variance of either operator is $\braket{\lambda_{\hat{Q}_{\text{leak}}}^{2}}_{\mathcal{S}_{\text{RB}}}=2$ for every state in $ \mathcal{S}_{\text{RB}}$, while $\braket{\lambda^{2}_{\hat{Q}_{\text{leak}}}}_{\mathcal{S}_{\text{leak},+}}=4$. This tells us the rate of population transfer from either state in $\mathcal{S}_{\text{RB}}$ to $\ket{\text{leak},+}$ is: 
\begin{eqnarray}
    \epsilon_{\text{leak}}&=& 2\!\!\!\sum_{k\in\mathcal{K}_{\text{leak}}}\!\!\!\gamma_{k}.
\end{eqnarray}
The rate of population transfer from $\mathcal{S}_{\text{leak}}$ back to $\mathcal{S}_{\text{RB}}$ is $2\epsilon_{\text{leak}}$. From symmetry, we know that when population leaks back from $\mathcal{S}_{\text{leak}}$ to $\mathcal{S}_{\text{RB}}$, it leaks equally to both states in $\mathcal{S}_{\text{RB}}$. The leakage component of the transfer matrix is then: 
\begin{equation}
    T_{\text{leak}} = \begin{bmatrix}
        -\epsilon_{\text{leak}} & 0 & \epsilon_{\text{leak}}  \\
        0 & -\epsilon_{\text{leak}} & \epsilon_{\text{leak}}  \\
        \epsilon_{\text{leak}} & \epsilon_{\text{leak}}  & -2\epsilon_{\text{leak}}  \\
    \end{bmatrix}. \\
\end{equation}
Together, this makes the total transfer matrix: 
\begin{eqnarray}\label{eq:tot_transfer_matrix}
    T &=& T_{\text{RB}}+T_{\text{leak}}.
\end{eqnarray}
As shown in the Supplementary Information, the general solution to this set of coupled differential equations is: 
\begin{align}
\label{eq_slerb_decay}
P_\textrm{survival} &= \frac{1}{3} + \frac{1}{2}(1-2\epsilon_{\textrm{\text{RB}}}-\epsilon_\textrm{leak})^{l} + \frac{1}{6}(1-3\epsilon_\textrm{leak})^{l}, \nonumber\\
P_\textrm{leak} &= \frac{1}{3} - \frac{1}{3}(1-3\epsilon_\textrm{leak})^{l}, \nonumber\\
P_\textrm{flip} &=  \frac{1}{3} - \frac{1}{2}(1-2\epsilon_\textrm{RB}-\epsilon_\textrm{leak})^{l} + \frac{1}{6}(1-3\epsilon_\textrm{leak})^{l}. \nonumber\\
\end{align}
Eq.~\ref{eq_slerb_decay} is the general analytical expression used for fitting population decays in the absence of SPAM errors. When non-negligible SPAM errors are present, a more general expression derived in the Supplementary Information (Eq.~\ref{eq:slrb_decay_spam}) can be used.
\subsubsection{Estimating average gate fidelities}

Eq.~\ref{eq:prob_not_transfer} gives the average probability of generating target state $\ket{g_{j,l+1}}=\hat{C}_{l+1}\ket{\psi_{j,l}}$, given some noise source $k$ and distribution of initial states $\mathcal{D}\equiv \{(\ket{\psi_{j,l}},P_{j,l})\}$. Defining the error channel $k$ as:
\begin{eqnarray}
    \mathcal{E}_{k}(\ket{\psi})\equiv \sum_{m,m^{\prime}}P_{m}\bra{m^{\prime}}\hat{U}_{e,k}(\ket{\psi}\bra{\psi}\otimes\ket{m}\bra{m})\hat{U}_{e,k}^{\dagger}\ket{m^{\prime}}, \nonumber \\
\end{eqnarray}
encapsulates the dependence on the state of the external degrees of freedom. If $\mathcal{D}$ forms a two-design on $\mathcal{S}_{2Q}$, we can replace the sum over $\mathcal{D}$ in Eq.~\ref{eq:prob_not_transfer} with a Haar measure, after which the equation reduces to:
\begin{eqnarray}
    \mathcal{F}_{g,k}\equiv \int d\psi\braket{\psi |\mathcal{E}_{k}(\ket{\psi})|\psi},
\end{eqnarray}
equal to the average gate fidelity in the presence of noise mechanism $k$, averaged over $\mathcal{S}_{2Q}$. If we sum over $\mathcal{I}_{g,k}\equiv 1-\mathcal{F}_{g,k}$ for every $k\in\mathcal{K}_{2Q}$, we get the total average gate infidelity:
\begin{eqnarray}\label{eq:avg_gate_infidelity}
    \mathcal{I}_{g} &\simeq & \sum_{k\in\mathcal{K}_{\text{RB}}}\braket{\lambda^{2}_{\hat{Q}_{k}}}_{\mathcal{S}_{2Q}}|\alpha_{k}|^{2}\sum_{m}P_{m}\braket{m |\hat{M}^{2}_{k}|m} \\
    &&+ \sum_{k\in\mathcal{K}_{\text{leak}}}\braket{\lambda^{2}_{\hat{Q}_{k}}}_{\mathcal{S}_{2Q}}|\alpha_{k}|^{2}\sum_{m}P_{m}\braket{m |\hat{M}^{2}_{k}|m}, \nonumber
\end{eqnarray}
where we have divided $\mathcal{K}$ into $\mathcal{K}_{\text{RB}}$ and $\mathcal{K}_{\text{leak}}$. Importantly, both terms are identical to their definitions in $\epsilon_{\text{RB}(\text{leak})}$\textemdash up to their dependence on $\mathcal{D}$, encapsulated entirely in the values of $\braket{\lambda^{2}_{\hat{Q}_{k}}}_{\mathcal{D}}$. It follows that we can estimate the average fidelity over $\mathcal{S}_{2Q}$ by adding $\epsilon_{\text{RB}}$ and $\epsilon_{\text{leak}}$, weighting them by the relative value of $\braket{\lambda^{2}_{\hat{Q}_{k}}}$ during SLERB to the same value, averaged instead over $\mathcal{S}_{2Q}$. Crucially, every operator in $\mathcal{Q}_{\text{leak}}$ has the same $\braket{\lambda^{2}_{\hat{Q}_{\text{leak}}}}_{\mathcal{S}_{\text{RB}}}=2$ and the same $\braket{\lambda^{2}_{\hat{Q}_{\text{leak}}}}_{\mathcal{S}_{2Q}}=8/5$. To estimate how much $\mathcal{K}_{\text{leak}}$ contributes to the total error budget, we weight $\epsilon_{\text{leak}}$ by:
\begin{eqnarray}\label{eq:ratio_var_rb}    \frac{\braket{\lambda^{2}_{\hat{Q}_{\text{leak}}}}_{\mathcal{S}_{2Q}}}{\braket{\lambda^{2}_{\hat{Q}_{\text{leak}}}}_{\mathcal{S}_{\text{RB}}}}&=&\frac{4}{5}.
\end{eqnarray}
The symmetric Pauli operators in $\mathcal{K}_{\text{RB}}$, either have $\braket{\lambda^{2}_{\hat{Q}_{\text{RB}}}}_{\mathcal{S}_{\text{RB}}}=2/3$ or $\braket{\lambda^{2}_{\hat{Q}_{\text{RB}}}}_{\mathcal{S}_{\text{RB}}}=0$. A SLERB experiment only measures errors in the former category, all of which are isometric with the single-qubit Pauli operators on $\mathcal{S}_{\text{RB}}$. The operators in $\mathcal{Q}_{\text{RB}}$ that act as single-qubit $\hat{\sigma}_{\phi}$ both have the same $\mathcal{S}_{2Q}$ averaged variance $\braket{\lambda^{2}_{\hat{Q}}}_{\mathcal{S}_{2Q}}=4/5$, while $\frac{1}{2}(\hat{\sigma}_{z,1}+\hat{z}_{z,2})|_{\mathcal{S}_{\text{RB}}}=\hat{\sigma}_{z}$ has an average variance of $\braket{\lambda^{2}_{\frac{1}{2}(\hat{\sigma}_{z,1}+\hat{\sigma_{z,2}})}}_{\mathcal{S}_{2Q}}=2/5$. To estimate the contribution of $\mathcal{K}_{\text{RB}}$ to $\mathcal{I}_{g}$, we take the larger of these two values $4/5$ and, again, weight $\epsilon_{\text{RB}}$ by ratio of the variance over $\mathcal{S}_{2Q}$ to the variance over $\mathcal{S}_{\text{RB}}$:
\begin{eqnarray}\label{eq:ratio_var_rb}
    \frac{\braket{\lambda^{2}_{\hat{Q}_{\text{RB}}}}_{\mathcal{S}_{2Q}}}{\braket{\lambda^{2}_{\hat{Q}_{\text{RB}}}}_{\mathcal{S}_{\text{RB}}}}&=&\frac{6}{5}.
\end{eqnarray}
This choice means $\hat{Q}_{\text{RB},k}= \frac{1}{2}(\hat{\sigma}_{z,1}+\hat{\sigma}_{z,2})$ errors will be double-counted in SLERB, potentially leading to an over-estimation of $\mathcal{I}_{g}$. Together, this makes the SLERB-estimated average Clifford infidelity:
\begin{eqnarray}\label{eq:slerb_avg_fidel}
    \mathcal{I}_{\text{Clifford}}&=&\frac{6}{5}\epsilon_{\text{RB}}+\frac{4}{5}\epsilon_{\text{leak}},
\end{eqnarray}
or, dividing by $13/6$ \textemdash the average number of $\hat{U}_{2Q}$ per Clifford\textemdash we get:
\begin{eqnarray}\label{eq:slerb_tq_fidel}
    \epsilon_\textrm{2Q}&=& \frac{6}{13}\mathcal{I}_{\text{Clifford}}
\end{eqnarray}
as our estimated average gate error.

\subsection{Group-theoretical properties}\label{sec:group_theory}

We now consider process matrix representations of the benchmarking group $\mathcal{G}_{\text{RB}}$ generated by the sampling procedure. We derive the irreducible representations (irreps) of the benchmarking group, which correspond to invariant subspaces under the action of the group. This allows us to connect the effects of twirling over $\mathcal{G}_{\text{RB}}$ to the exponential decay of population in and between the irreps. Our experiment measures some of these decays and leaves others unmeasured. A \textit{consistent} estimator \footnote{A consistent estimator is one that converges in probability to the true value as the number of data points is increased to infinity} of gate fidelity would necessarily need to measure all these decays. Nevertheless, we average the measured decays to extrapolate the unmeasured decays, which produces an approximate fidelity estimator. This fidelity estimator makes different assumptions than the one in Eq. (\ref{eq:slerb_avg_fidel}) yet yields remarkable agreement. Furthermore, our group-theoretical analysis reveals that some of the assumptions of the prior section are enforced by twirling over $\mathcal{G}_{\text{RB}}$. 

Our protocol deploys a direct generalization of single-qubit RB with leakage on an invariant subspace of the MS gate. To establish the relationship, we begin by considering the unitaries that are generated by setting the MS phase to 0 and $\pi/4$: 
\begin{eqnarray}\label{eq:msx_msy}
    \hat{M}_x &\equiv \hat{U}_{2Q}(\tfrac{\pi}{2},0),  \\
    \hat{M}_y &\equiv \hat{U}_{2Q}(\tfrac{\pi}{2},\tfrac{\pi}{4}).
\end{eqnarray}
It can be shown (see Supplementary Information) that the target action of these gates is isomorphic to a block-diagonal decomposition of the action on the even-parity subspace $\mathcal{S}_{\text{RB}}$ and the odd-parity subspace $\mathcal{S}_\text{leak}$
\begin{eqnarray}\label{eq:msx_msy}
    \hat{M}_x &\cong R_x^{00, 11}(\tfrac{\pi}{2}) \oplus R_x^{01, 10}(\tfrac{\pi}{2}), \\
    \hat{M}_y &\cong R_y^{00, 11}(\tfrac{\pi}{2}) \oplus R_x^{01, 10}(\tfrac{\pi}{2}),
\end{eqnarray}
where $R_x^{a, b}(\theta_{\text{\text{MS}}})$ and $R_y^{a, b}(\theta_{\text{\text{MS}}})$ are generalized SU(2) rotations about $x$ and $y$ over the subspace spanned by the states $\ket{a}, \ket{b}$. The benchmarking group generated by $\hat{M}_x$ and $\hat{M}_y$ acts on the even-parity subspace $\mathcal{S}_{\text{RB}}$ in the same way that the $x$ and $y$ Pauli rotations of single-qubit RB act on the associated qubit subspace. Thus, our gate set forms a two-design over this subspace, and any error sources that preserve $\mathcal{S}_{\text{RB}}$ and do not commute with the benchmarking group are completely depolarized on this subspace. However, the action of the benchmarking group on the odd-parity subspace $\mathcal{S}_{\text{leak}}$ is restricted to rotations about the $x$-axis; this does not completely depolarize noise on this subspace and complicates the irrep structure of the representation of the benchmarking group. 

\subsubsection{Decay model}

To derive the expected decay forms for the benchmarking protocol, we begin by identifying the benchmarking group generated by $\hat{M}_x$ and $\hat{M}_y$ 
\begin{equation}
    \mathcal{G}_{RB} = \langle \hat{M}_x, \hat{M}_y \rangle.
\end{equation}
This group has 96 elements, and as can be seen from the character table, Tab. \ref{tab:char_table}, is not a separable product group of uncoupled action on $\mathcal{S}_\text{RB}$ and $\mathcal{S}_\text{leak}$. We consider a representation $V$ of the group $\mathcal{G}_\text{RB}$ as process matrices acting on the Hilbert-Schmidt space of two-qubit vectorized density matrices. 

Now, consider the twirl $\tilde{\Lambda}_{\mathcal{G}}$ of a noise channel $\Lambda$ over the group $\mathcal{G}_{RB}$: 
\begin{equation}\label{eq:twirl_def}
    \tilde{\Lambda}_{\mathcal{G}} = \frac{1}{|G|} \sum_{g \in \mathcal{G}_{RB} } V(g) \Lambda V(g)^{-1}. 
\end{equation}
Schur's lemma implies that the action of the twirled channel is isomorphic to a direct sum over actions on irreps of the representation~\cite{Claes_2021}
\begin{equation}
    \tilde{\Lambda}_\mathcal{G} \cong \bigoplus_{\nu} Q_\nu \otimes \mathbbm{1}_\nu, 
\end{equation}
where $\nu$ indexes the irreps. To derive the expected decay signal for the benchmarking experiment, one must calculate the overlaps between the irreps and the given state preparation and measurement pair used in the experiment. For an initial state preparation of $\ket{00}$ or $\ket{11}$, the only irreps that have non-zero overlap are the trivial irrep and the irrep corresponding to depolarization on $\mathcal{S}_{\text{RB}}$, see Tab. \ref{tab:overlaps} in the Supplementary Information; here, we show that the generic decay form of the success probability in the absence of SPAM errors consists of a constant offset and three exponential terms
\begin{equation}
    P_{\text{survival}} = c_1 + c_2 q_{\text{leak}, +}^l + c_3 q_{\text{leak}, -}^l + \frac{1}{2} q_{\text{RB}}^l.
\end{equation}
One exponential $q_{\text{RB}}^l$ corresponds to depolarization in $\mathcal{S}_{\text{RB}}$, while the other two correspond to leakage between this subspace and $\mathcal{S}_\textrm{leak} = \mathcal{S}_\textrm{leak, +} \cup \mathcal{S}_\textrm{leak, -}$. This is the most specific model one can derive without assumptions about the noise in the system. Assuming, as in Sec.~\ref{sec:transfer_matrix}, that the errors are symmetric, we can derive the expected decay model (see the Supplementary Information): 
\begin{equation}\label{eq:P_survival}
    P_\text{survival} = \frac{1}{3} + \frac{1}{2} q_{\text{RB}}^m + \frac{1}{6} q_{\text{leak}, +}^m,
\end{equation}
and similarly 
\begin{equation}\label{eq:P_flip}
    P_\text{flip} = \frac{1}{3} - \frac{1}{2} q_{\text{RB}}^m + \frac{1}{6}q_{\text{leak}, +}^l,
\end{equation}
where $q_{\text{RB}}$ is a decay term corresponding to errors in $\mathcal{S}_{\text{RB}}$ and $q_{\text{leak},+}$ is a decay term corresponding to leakage between $\mathcal{S}_{\text{RB}}$ and $\mathcal{S}_{\text{leak}, +}$. This decay form is essentially the same as Eq. (\ref{eq_slerb_decay}), derived from the transfer matrix model. 

\subsubsection{Extended fidelity estimator}

As discussed in the Supplementary Information, the effective fidelity can be calculated given the knowledge of all subspace decays $q_{\nu, \mu}$. However, as shown in the previous section, SLERB only directly measures $q_\text{RB}$ and $q_{\text{leak}, \pm}$. Our approach is therefore to report the decays on unmeasured subspaces by \textit{extending} the measured decays to approximate the decays on unmeasured subspaces. One way to do this is by setting $q_{\text{leak}, +} = q_{\text{leak}, -}$ (pessimistic for symmetric errors), and averaging the measured decays to approximate the unmeasured ones. While not applicable to arbitrary adversarial noise, our analysis indicates these are conservative approximations for the anticipated noise sources, and numerical simulations in Sec.~\ref{sec:sims} show that the estimator closely approximates the true error rate for random unitary error channels. As shown in the Supplementary Information, setting  all unmeasured terms to the uniform average of measured terms results in a process fidelity metric
\begin{equation}\label{eq:group_theory_channel_fidel}
    \mathcal{F} = \frac{1 + 8 q_{\text{RB}} + 7 q_{\text{leak}, +}}{16}, 
\end{equation}
which can be converted to an average Clifford gate fidelity~\cite{nielsen_2002} and reported as 
\begin{equation}\label{eq:group_theory_avg_fidel}
    \bar{\mathcal{F}} = \frac{5 + 8 q_{\text{RB}} + 7 q_{\text{leak}, +}}{20}. 
\end{equation}
We can further convert to an expression for infidelity in terms of $\epsilon_{\text{RB}}$ and $\epsilon_{\text{leak}}$ 
\begin{equation}\label{eq:slerb_group_theory_fidelity}
    \bar{\mathcal{I}}_\text{Clifford} = \frac{4}{5} \epsilon_\text{RB} + \frac{29}{20} \epsilon_\text{leak}.
\end{equation}
In practice, the Cliffords are compiled from native gate sequences. Thus, an ``error per gate'' metric would rescale the extended Clifford infidelity by the average number of gates per Clifford 
\begin{equation}
    \epsilon_\textrm{2Q} = \frac{6}{13}\bar{\mathcal{I}}_\text{Clifford}. 
\end{equation}

This estimator of Eq. (\ref{eq:group_theory_avg_fidel}) should be compared with Eq. (\ref{eq:slerb_avg_fidel}). While these two estimators are not exactly equivalent, the small discrepancy between the two can be explained by differing assumptions. Eq. (\ref{eq:slerb_avg_fidel}) is based on the assumptions of 1) incoherent population transfer, 2) leakage only within the symmetric subspace, 3) equal leakage and seepage rates, and 4) small errors. In contrast, Eq. (\ref{eq:group_theory_avg_fidel}) is derived by analyzing the decays on irreps that are enforced by twirling over $\mathcal{G}_{RB}$, and extending the measured decays to approximate unmeasured decays. Its assumptions are 1) twirling over $\mathcal{G}_{\text{RB}}$, 2) independent noise on each Clifford, 3) approximately unbiased noise. If one were to take a weighted, instead of uniform, average between the measured decays, then Eq. \ref{eq:slerb_avg_fidel} could be recovered for a certain weighting. 

The analysis in this section has shown that twirling over the benchmarking group $\mathcal{G}_{\text{RB}}$ enforces some of the assumptions made in the prior section. Namely, we have shown that twirling over $\mathcal{G}_\text{RB}$ enforces the assumption of incoherent population transfer within $\mathcal{S}_{\text{RB}}$ and to/from $\mathcal{S}_{\text{RB}}$ and $\mathcal{S}_{\text{leak}}$. The analysis has also shown that twirling over $\mathcal{G}_{\text{RB}}$ combined with assumptions of leakage only in the symmetric space and equal leakage/seepage rates enforces the decay forms of Eq. \ref{eq_slerb_decay}.

Our derivations in this section also provide the necessary framework to construct a consistent estimator of gate fidelity with extensions to the current protocol. As we have said, we measure only two of the relevant decays required to construct a consistent fidelity estimate. One can extend the current protocol to measure all the subspace decays by including more state preparation and measurement pairs in the protocol. For example, if one were to prepare in $\ket{01}$ or $\ket{10}$, then the resulting RB signal would be sensitive to error processes that cause transfer between $\mathcal{S}_{\text{leak}, +}$ and $\mathcal{S}_{\text{leak}, -}$. In this case, the signal may contain oscillations due to residual coherence from partial twirling, but the theory can still be applied. Similarly, preparing superposition states between $\mathcal{S}_{\text{leak}}$ and $\mathcal{S}_\text{RB}$ allows us to measure decays associated with forward and backward mixing processes (Tab. \ref{tab:irrep_properties}). Employing the mechanisms of character RB~\cite{helsen_new_2019}, filtered RB~\cite{Helsen_2022}, or synthetic SPAM RB~\cite{fan2024randomizedbenchmarkingsyntheticquantum} would provide a general framework to analyze the resulting signals and extract a consistent estimator of $\mathcal{F}$. 

\subsection{MS gate error symmetry}\label{sec:symmetry}

While SLERB is a highly general technique, the analytical expressions for populations derived in the previous sections are based on the assumption that errors keep the populations within the symmetric subspace $\{\ket{00}, \ket{11},\frac{1}{\sqrt{2}}(\ket{01} + \ket{10})\}$. This assumption is well motivated by the physics of MS interactions; as each qubit is coupled to the same motional mode, the populations stay within the symmetric subspace even if each qubit couples to the MS drive field with a different strength. As a result, only a handful of physically plausible error channels have the potential to populate the antisymmetric subspace, notably photon scattering and differential qubit frequency shifts. We expect both of these to be negligible for the MS gate implementation in Sec.~\ref{sec:implementation}, as the gate involves no laser fields (eliminating photon scattering), and uses dynamical decoupling (suppressing qubit frequency shifts). In Sec.~\ref{sec:full_ms_sims}, we verify this reasoning through direct numerical simulations of SLERB sequences with additional error sources.

Finally, we note that SLERB offers two natural ways of verifying the symmetry assumption. The first is through the measurement of population asymptotes $\vec{P}_{l\rightarrow \infty}$. Eq.~\ref{eq_slerb_decay} shows that for $l\rightarrow \infty$, if $\epsilon_\text{leak}>0$, the populations will reach an equilibrium state of ${\vec{P}_{l\rightarrow \infty}\rightarrow (1/3,1/3,1/3)}$. On the other hand, errors that drive population to $\frac{1}{\sqrt{2}}(\ket{01} - \ket{10})$ will result in ${\vec{P}_{l\rightarrow \infty}\rightarrow (1/4,1/4,1/2)}$. Thus, the presence of any such error would result in a measurable change in the asymptotes. Second, SLERB can measure such errors directly by modifying the sequence to act on the $\ket{01}, \ket{10}$ subspace instead. However, this subspace would require asymmetric MS interactions where $\phi_\textrm{MS}$ is different for each ion, see Supplementary Information. 
\section{Numerical simulations}
\label{sec:sims}

We have performed extensive simulations of the protocol to verify the accuracy and precision of our fidelity estimator. These simulations have been written at three levels of abstraction: 1) state vector simulation of Clifford sequences with unitary errors acting on pure initial states, 2) direct simulation of the physical noise on the full MS dynamics that include both qubit and motional degrees of freedom, and 3) direct calculation of the twirled channel, which represents the infinite shot and randomization limit. Additionally, we use the full MS simulations to investigate the effect of common MS errors on the population dynamics for both the symmetric and antisymmetric subspaces.

\subsection{Population transfer rates}

\begin{figure}[!h]
\includegraphics[width=1\columnwidth]{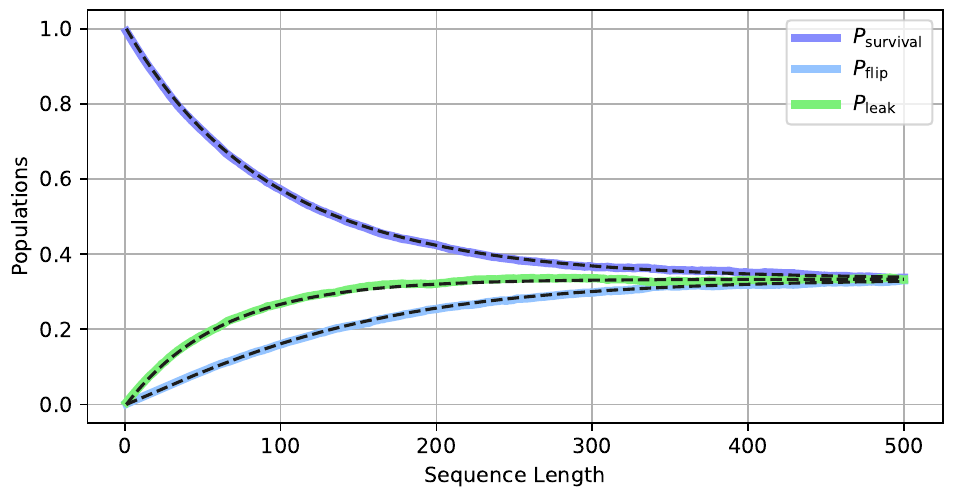}
\centering
\caption{\label{fig:clifford_avg} Monte Carlo simulations (colored solid lines) of population transfer for a SLERB sequence of length $l=500$, averaging over $5\times10^{3}$ random circuits of single-qubit Cliffords on $\mathcal{S}_{\text{RB}}$, all of which decompose into Eq.~(\ref{eq:ideal_unitary}). For this run, we multiplied each Clifford by error unitaries representing two noise unitaries: one in $\mathcal{K}_{\text{RB}}$ and one in $\mathcal{K}_{\text{leak}}$. Using the results of Sec.~\ref{sec:transfer_matrix}, we calculate $\epsilon_{\text{RB}}$
and $\epsilon_{\text{leak}}$ analytically, then apply the resultant transfer matrix $T$ to the initial population vector $\vec{P}_{0}=(1,0,0)$ (black dashed lines). The result illustrates how the analytical expressions agree with numerical simulations.}
\end{figure}

To verify the analytical transfer matrix results in Sec.~\ref{sec:transfer_matrix}, we use a Monte Carlo simulation that averages over random $\mathcal{S}_{\text{RB}}$ single-qubit Cliffords. Each Clifford is generated by numerically iterating through decompositions of Eq.~(\ref{eq:ideal_unitary}) until we obtain $24$ unique single-qubit Cliffords on $\mathcal{S}_{\text{RB}}$. We apply two error unitaries:
\begin{eqnarray}
    \hat{U}_{\text{RB},e}&=& e^{-i\alpha_{\text{RB}}\hat{\sigma}_{x,1}\hat{\sigma}_{x,2}},
\end{eqnarray}
as an example of an error that conserves $\mathcal{S}_{\text{RB}}$, and: 
\begin{eqnarray}
    \hat{U}_{\text{leak},e}&=& e^{-i\alpha_{\text{leak}}(\hat{\sigma}_{x,1}+\hat{\sigma}_{x,2})},
\end{eqnarray}
as a leakage error; to ensure our assumption that leakage is incoherent, we apply a random sign to $\pm\alpha_{\text{leak}}$ at every occurrence of $\hat{U}_{\text{leak},e}$. So if $\hat{C}_{l}$ is the $l^{\text{th}}$ Clifford of a sequence, the unitary applied in the simulation is:
\begin{eqnarray}
    \hat{U}_{t,l}=\hat{U}_{\text{2q},e}\hat{U}_{\text{leak},e}\hat{C}_{l}.
\end{eqnarray}
The unitary $\hat{U}_{\text{RB},e}$ represents a noise source in $\mathcal{K}_{\text{RB}}$, and the resulting population transfer rate within $\mathcal{S}_{\text{RB}}$ is:
\begin{eqnarray}
    \epsilon_{\text{RB}} = \frac{2}{3}|\alpha_{\text{RB}}|^{2},
\end{eqnarray}
as discussed in Sec.~\ref{sec:transfer_matrix}. The unitary $\hat{U}_{\text{leak},e}$ is in $\mathcal{K}_{\text{leak}}$, and gives a leakage rate:
\begin{eqnarray}
    \epsilon_{\text{leak}} = 2|\alpha_{\text{leak}}|^{2}.
\end{eqnarray}
For the example in Fig.~~\ref{fig:clifford_avg}, we set $\alpha_{\text{RB}}=\alpha_{\text{leak}}=\pi/60$. Plugging these into our transfer matrix $T$, we show the results from our Monte-Carlo simulation for systems initialized to $\vec{P}_{0}=(1,0,0)$ and averaged over $5\times10^{3}$ random trials of circuits up to length $l=500$. We compare this to the transfer matrix model $\vec{P}_{l}=T^{l}\vec{P}_{0}$. The figure demonstrates that the two calculations converge, with no fitting parameters.

\subsection{Full MS dynamics}
\label{sec:full_ms_sims}

We also perform full Hamiltonian-level simulations of SLERB sequences in QuTiP~\cite{johansson_2013}. For each MS gate, we model the interaction as

\begin{align}
\label{eq:ms_dynamics}
    \hat{H}_\textrm{MS} = \frac{\hbar\Omega_\textrm{MS}}{2}\sum_{n={1,2}}&\left[\hat{\sigma}_{x, n}{\cos\phi_{\textrm{MS}}} + \hat{\sigma}_{y, n}{\sin\phi_{\textrm{MS}}}\right] \\ \nonumber
    &(\hat{a}e^{i\delta t}+\hat{a}^\dagger e^{-i\delta t}),
\end{align}

\noindent where $\hat{\sigma}_{i,n}$ is the Pauli operator for qubit $n$. The strength of the spin-motion coupling is described by $\Omega_{\textrm{MS}}$, and $\delta$ is the detuning of the interaction from the motional mode of the trapped ions, described by the annihilation and creation operators $\hat{a}$ and $\hat{a}^\dagger$, respectively. We assume the ions' spins only couple to a single mode of motion; the initial state is $\ket{00}$, and the ground state of the motional mode. We model a two-loop Walsh-modulated gate by changing $\phi_\textrm{MS}$ by $\pi$ halfway through the gate~\cite{hayes_2012}. We set $\Omega_\textrm{MS}/2\pi=1$, and thus $t_g=1/\sqrt{2}$, and $\delta/2\pi=2\sqrt{2}$. We generate the SLERB sequences in the exact same way as the experimental implementation (Sec.~\ref{sec:implementation}), using 50 random circuits for each sequence length.

\begin{figure}[!h]
\includegraphics[width=1\columnwidth]{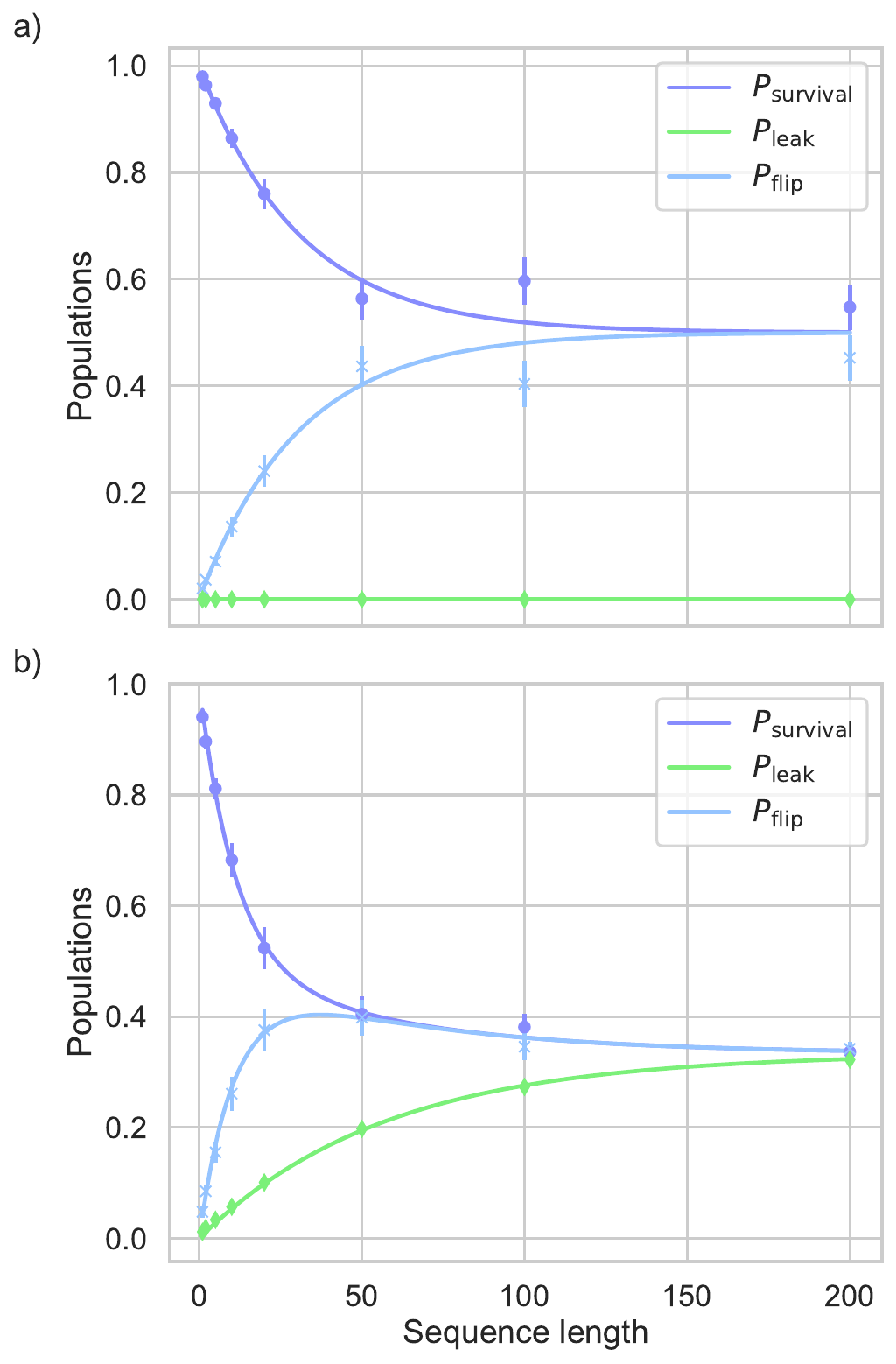}
\centering
\caption{\label{fig_ms_sim} Numerical simulations of SLERB sequences for (a) amplitude and (b) detuning errors. For each sequence length, we sample 50 random sequences; the error bars correspond to the standard deviation for each population from the 50 circuits. The solid lines correspond to fits to the data following Eq.~\ref{eq_slerb_decay}. (a) For amplitude errors, there is no leakage and the states stay within the $\ket{00}, \ket{11}$ subspace, with asymptotes of 1/2 for $P_\textrm{survival}$ and $P_\textrm{leak}$. (b) For detuning errors, there is residual spin-motional entanglement, which results in leakage. The asymptotes are instead $1/3$ for all three populations.
}
\end{figure}

We first investigate the effect of the gate Rabi frequency offset by setting $\Omega_\textrm{MS}/2\pi=(1+0.05)$. Fig.~\ref{fig_ms_sim}a) shows that, as expected for an error that only affects the geometric phase, there is no leakage. Thus, $P_\textrm{leak}=0$, and the other two populations have an asymptote at $1/2$. Fig.~\ref{fig_ms_sim}b) instead presents simulation of a motional mode frequency offset $\delta/2\pi=(1-0.07)\times2\sqrt{2}$. In contrast to the Rabi frequency offset, this error results in leakage as there is residual spin-motion entanglement at the end of each gate. Here, the asymptotes are instead 1/3 for all three populations, as all states within the symmetric subspace are populated equally. We also observe the multi-exponential decay as we expect from Eq.~\ref{eq_slerb_decay}. In the simulations, we reset the mode to the ground state to prevent any coherent buildup of errors. We discuss in the Supplementary Information the behavior when the motional mode is not reset.

Additionally, we perform individual simulations of common error mechanisms in laser-free MS gates to verify the error symmetry assumption in Sec.~\ref{sec:symmetry}. The results are summarized in Table~\ref{table_errors}. We find that all global errors -- i.e., when both qubits experience an identical offset in a control parameter -- keep populations within the symmetric subspace. Further, errors that only effect the geometric phase, but do not cause any residual spin-motion entanglement, keep populations within the $\ket{00}, \ket{11}$ subspace. This behaviour holds even when each ion has a different spin-motion coupling strength. Motional errors, such as mode frequency offsets, motional dephasing, or motional heating, can additionally populate the $\frac{1}{\sqrt{2}}(\ket{01}+\ket{10})$ state but still keep populations within the symmetric subspace. One exception is a differential qubit frequency shift, which populates the antisymmetric state $\frac{1}{\sqrt{2}}(\ket{01}-\ket{10})$, but only if there is leakage to the $\frac{1}{\sqrt{2}}(\ket{01}+\ket{10})$ to begin with, making it a second-order effect. Similarly, if an additional carrier field is used for dynamical decoupling (see Sec.~\ref{sec:implementation}), a differential carrier Rabi frequency only results in population to the singlet state if the carrier does not commute with another interaction, such as the spin-dependent force or a global qubit frequency offset.

\begin{table}\centering
\begin{tabular}{p{3cm} c c c c}  \toprule
\multirow{3}{2em}{\textbf{Errors}} & $\ket{00}$ & $\ket{00}$, $\ket{11}$ & $\ket{00}$, $\ket{11}$ & $\ket{01} + \ket{10}$ \\ 
& $\Updownarrow$ & $\Updownarrow$ & $\Updownarrow$ &$\Updownarrow$ \\
& $\ket{11}$ & $\ket{01} + \ket{10}$ & $\ket{01} - \ket{10}$ & $\ket{01} - \ket{10}$ \\
\midrule
\multirow{2}{3cm}{Global qubit frequency offset} & \multirow{2}{*}{\Checkmark} & \multirow{2}{*}{\Checkmark} & \multirow{2}{*}{\XSolidBrush} & \multirow{2}{*}{\XSolidBrush}  \\
\\ 
\multirow{1}{*}{Motional errors} & \Checkmark & \Checkmark & \XSolidBrush & \XSolidBrush \\ 
Global $\Omega_\textrm{MS}$ offset & \Checkmark & \XSolidBrush & \XSolidBrush & \XSolidBrush \\ 
Differential $\Omega_\textrm{MS}$& \Checkmark & \XSolidBrush & \XSolidBrush & \XSolidBrush  \\ 
\multirow{4}{3cm}{Global carrier phase offset, identical carrier Rabi frequencies} & \multirow{4}{*}{\Checkmark} & \multirow{4}{*}{\Checkmark} & \multirow{4}{*}{\XSolidBrush} & \multirow{4}{*}{\XSolidBrush}  \\ \\ \\ \\
\multirow{2}{3cm}{Differential qubit frequency shift}& \multirow{2}{*}{\Checkmark}   &  \multirow{2}{*}{\XSolidBrush}& \multirow{2}{*}{\XSolidBrush} & \multirow{2}{*}{\Checkmark}  \\ \\
\multirow{4}{3cm}{Global carrier phase offset, differential carrier Rabi frequencies} & \multirow{4}{*}{\Checkmark} & \multirow{4}{*}{\Checkmark} & \multirow{4}{*}{\Checkmark} & \multirow{4}{*}{\Checkmark}  \\ \\ \\ \\
\bottomrule
\end{tabular}
\caption{\label{table_errors} Population transfer dynamics of different laser-free MS gate errors. For each error, we mark the allowed population transfers with \Checkmark, and those not allowed with \XSolidBrush. In general, geometric phase errors keep the states within the $\{\ket{00},\ket{11}\}$ subspace, while any motional errors result in populations of the $\ket{01}+\ket{10}$ state as well. Antisymmetric errors can result in populations in the $\ket{10}-\ket{10}$ state. However, the differential qubit frequency shifts only result in populations in the antisymmetric state if there is some leakage to begin with. In the absence of any leakage, differential qubit frequency shifts keep the populations within the $\ket{00}$, $\ket{11}$ subspace. Similarly, differential carrier Rabi frequencies only populate the antisymmetric state when the carrier does not commute with the gate, which causes leakage.}
\end{table}

\subsection{Random unitary error channels}

Finally, we verify that our protocol performs well for random unitary noise even without the assumption of symmetric noise. These simulations were performed via direct calculation of the twirled process matrix associated with a unitary error. Explicitly, we construct a random unitary error $\hat{E}$ via 
\[
    \hat{E} = \exp\left( i \vec{\theta} \cdot \bm{P} \right),
\]
where $\vec{\theta}$ is a random variable and $\bm{P}$ is a basis of two-qubit Pauli matrices. We choose the distribution of $\vec{\theta}$ to be a multivariate Gaussian with uniform variance equal to $0.01$. This choice of distribution results in an expected fidelity in the range observed experimentally in Sec.~\ref{sec:implementation}. We then convert each unitary error into a process matrix representation $\Lambda = \hat{E}^* \otimes \hat{E}$ and calculate the twirled channel $\tilde{\Lambda}_{\mathcal{G}}$ via Eq. (\ref{eq:twirl_def}), where $\mathcal{G}_{\text{RB}}$ is the entire benchmarking group of 96 elements. Calculation of the twirled channel requires averaging over the entire benchmarking group rather than a subset of 24 elements. In this way, we directly simulate the theoretical limit in the case of infinite shots and all randomizations. We then calculate $P_\text{survival}$, $P_\text{flip}$, and $P_\text{leak}$ for varying $l$. Finally, we fit the resulting exponential decays to the model 
\begin{equation}
    P_\text{survival} + P_\text{flip} = A + B q_\text{\text{leak}, +}^l\
\end{equation}
and 
\begin{equation}
    P_\text{survival} - P_\text{flip} = C q_\text{RB}^l.
\end{equation}
Note that the unitary simulations in this subsection do not respect the symmetric leakage assumption that has been made in the quantitative analysis in Sec.~\ref{sec:theory} and validated experimentally in Sec.~\ref{sec:implementation}. Hence, we would expect to see a second exponential term in $P_\text{survival} + P_\text{flip}$. However, we find that the symmetric model still provides a good estimate of gate fidelity, i.e., that fitting a single-exponential model to a double exponential decay still captures the dominant decay profile, resulting in relatively small approximation error. 

\begin{figure}
    \centering
    \includegraphics[width=\linewidth]{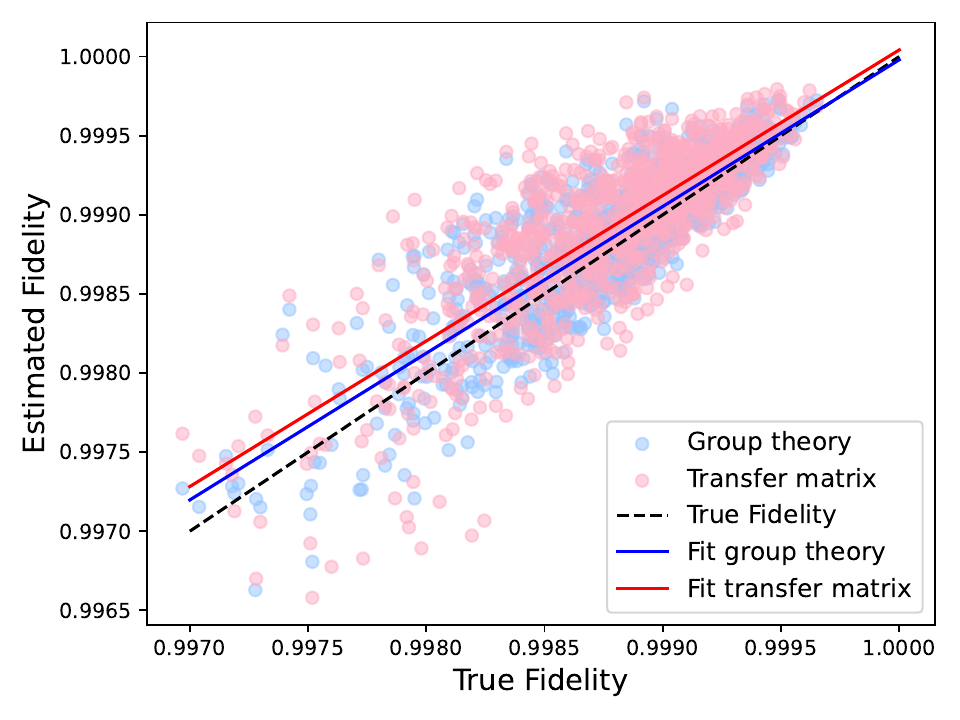}
    \caption{Estimated vs true channel fidelity for 1000 random unitary error channels. Each point corresponds to a full simulation of a benchmarking experiment where a random unitary error channel was sampled, the numeric twirl of the resulting channel representation was calculated, and the signals were calculated and fit to extract an estimate of the fidelity. Transfer matrix points correspond to the fidelity estimator of Eq. \ref{eq:slerb_avg_fidel}, and group theory points correspond to Eq. \ref{eq:group_theory_avg_fidel}. The lines represent fits of the two estimated distributions to a linear model. }
    \label{fig:random_unitary_errors_plot}
\end{figure}

We finally estimate the fidelity using Eqs. \ref{eq:slerb_avg_fidel} and \ref{eq:group_theory_avg_fidel}. We repeated this protocol 1000 times. The results are plotted in Fig. \ref{fig:random_unitary_errors_plot}. We further calculate the relative error $e_\mathcal{F}$ between the true channel fidelity $\tilde{\mathcal{F}}$ and the estimated fidelity $\hat{\mathcal{F}}$:
\begin{equation}
    e_\mathcal{F} = \frac{|\hat{\mathcal{F}} - \tilde{\mathcal{F}}|}{\tilde{\mathcal{F}}}. 
\end{equation}
We find that, for random unitary error channels in the simulated fidelity regime, the average relative error between each estimator and the truth is $16 \pm 12 \%$ for the group theory estimator and $22 \pm 15 \%$ for the transfer matrix estimator.

\section{Experimental implementation}
\label{sec:implementation}
We implement SLERB in a cryogenic ion trap setup similar to that described in Ref.~\cite{loeschnauer_2024}. Qubits are encoded in the $4S_{1/2}$ Zeeman sublevels of $\mathrm{^{40}Ca^+}$ ions, each with a qubit frequency of $\omega_0 \approx 2 \pi \times 240$~MHz. The MS gates are performed on the in-plane out-of-phase radial mode of a two-ion crystal at $\omega_m \approx 2 \pi \times 3.5$~MHz by passing a bichromatic oscillating current $\omega_0 \pm (\omega_m + \delta)$ through an on-chip antenna which generates magnetic field gradients~\cite{wineland_1998, mintert_2001, ospelkaus_2008}. An additional carrier tone at $\omega_0$ dynamically decouples from errors due to qubit frequency fluctuations~\cite{harty_2016}. 

\begin{figure}[!h]
\includegraphics[width=1\columnwidth]{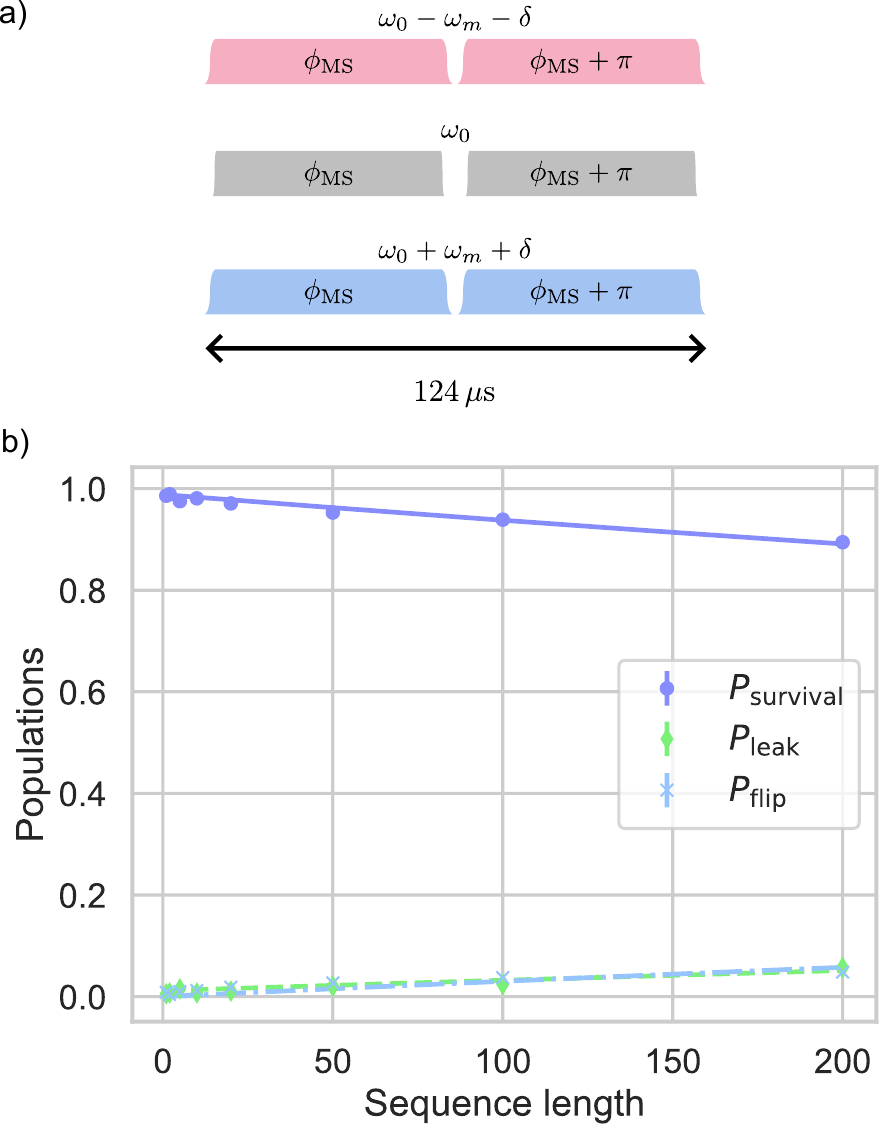}
\centering
\caption{\label{fig_example} SLERB implementation and results. (a) Pulse sequence for an individual MS gate with phase $\phi_\textrm{MS}$. Each MS gate consists of three frequencies: blue and red sideband tones at $\omega_0 \pm(\omega_m+\delta)$, and a carrier at $\omega_0$ for dynamical decoupling. We set the phases of all the tones to $\phi_\textrm{MS}$ in the first half of the gate, and to $\phi_\textrm{MS} + \pi$ in the second half of the gate. (b) SLERB results with up to 200 Cliffords. The populations $P_\textrm{survival}$ (purple circles), $P_\textrm{flip}$ (blue crosses), and $P_\textrm{leak}$ (green diamonds) indicate the populations in the target state, in the incorrect state within the $\{\ket{00},\ket{11}\}$ subspace, and the population that has leaked out of this subspace into the $\frac{1}{\sqrt{2}}(\ket{01}+\ket{10})$ state, respectively. The measurements are performed with 50 randomizations and 50 shots each; error bars indicate the standard deviation of the populations across all the randomizations for each sequence length.}
\end{figure}

The pulse sequence for an MS gate with phase $\phi_{\mathrm{MS}}$ is shown in Fig.~\ref{fig_example}a). The sideband tones lead to a gate interaction with strength $\Omega_{\mathrm{MS}} \approx 2 \pi \times 6$~kHz, detuning $\delta \approx 2 \pi \times 17$~kHz, and on/off ramp duration of $\approx 3$~$\mu$s. The phase $\phi_{\mathrm{MS}}$ of these tones is incremented by $\pi$ halfway through the gate to achieve Walsh-1 modulation~\cite{hayes_2012}, reducing errors due to offsets in the gate mode frequency. The carrier tone has Rabi frequency $\Omega_c \approx 2\pi \times 40$~kHz and on/off ramp duration of $\approx 0.5$~$\mu$s. The phase of the carrier tone is updated by $\pi$ halfway through the gate to ensure minimal residual carrier rotation by the end of the gate. The total duration of the gate is $124\,\mu$s.

The experiment is implemented as follows. At the start of every shot, the ions are initialized in $\ket{00}$ and cooled to close to the ground state in all 6 modes of motion, with an average phonon occupation $\bar{n}=0.05(1)$ in the gate mode. We then perform a SLERB sequence, followed by a measurement in the computational basis. The target state of the sequence, either $\ket{00}$ or $\ket{11}$ with Pauli-frame randomisation, is assigned to $P_\textrm{survival}$, while the population in the other state within this subspace is $P_\textrm{flip}$. The population $P_\textrm{leak}$ corresponds to measurements of either $\ket{01}$ or $\ket{10}$; our detection does not distinguish between $\ket{01}+\ket{10}$ or $\ket{01}-\ket{10}$. 

    Figure~\ref{fig_example}(b) shows the results for a SLERB sequence with up to 200 Cliffords (on average 433 two-qubit gates). We fit the data to Eq.~\ref{eq:slrb_decay_spam}, which modifies Eq.~\ref{eq_slerb_decay} to include SPAM errors, finding ${\epsilon_\textrm{RB}=3.2(3)\times10^{-4}}$, ${\epsilon_\textrm{leak}=2.2(3)\times10^{-4}}$, ${\bar{\epsilon}_\textrm{SPAM}=5.9(6)\times10^{-3}}$ and finally a two-qubit gate error ${\epsilon_\textrm{2Q}=2.6(2)\times10^{-4}}$, following Eq.~\ref{eq:slerb_avg_fidel}. Using Eq.~\ref{eq:slerb_group_theory_fidelity} instead, we obtain ${\epsilon_\textrm{2Q}=2.7(2)\times10^{-4}}$, which is statistically indistinguishable. The error bars correspond to the 68\% confidence interval from non-parametric bootstrapping with 10,000 samples. These results are consistent with the error obtained by our team in Ref.~\cite{loeschnauer_2024} using partial state tomography.

\subsection{Calibrations}

\begin{figure}[!h]
\includegraphics[width=1\columnwidth]{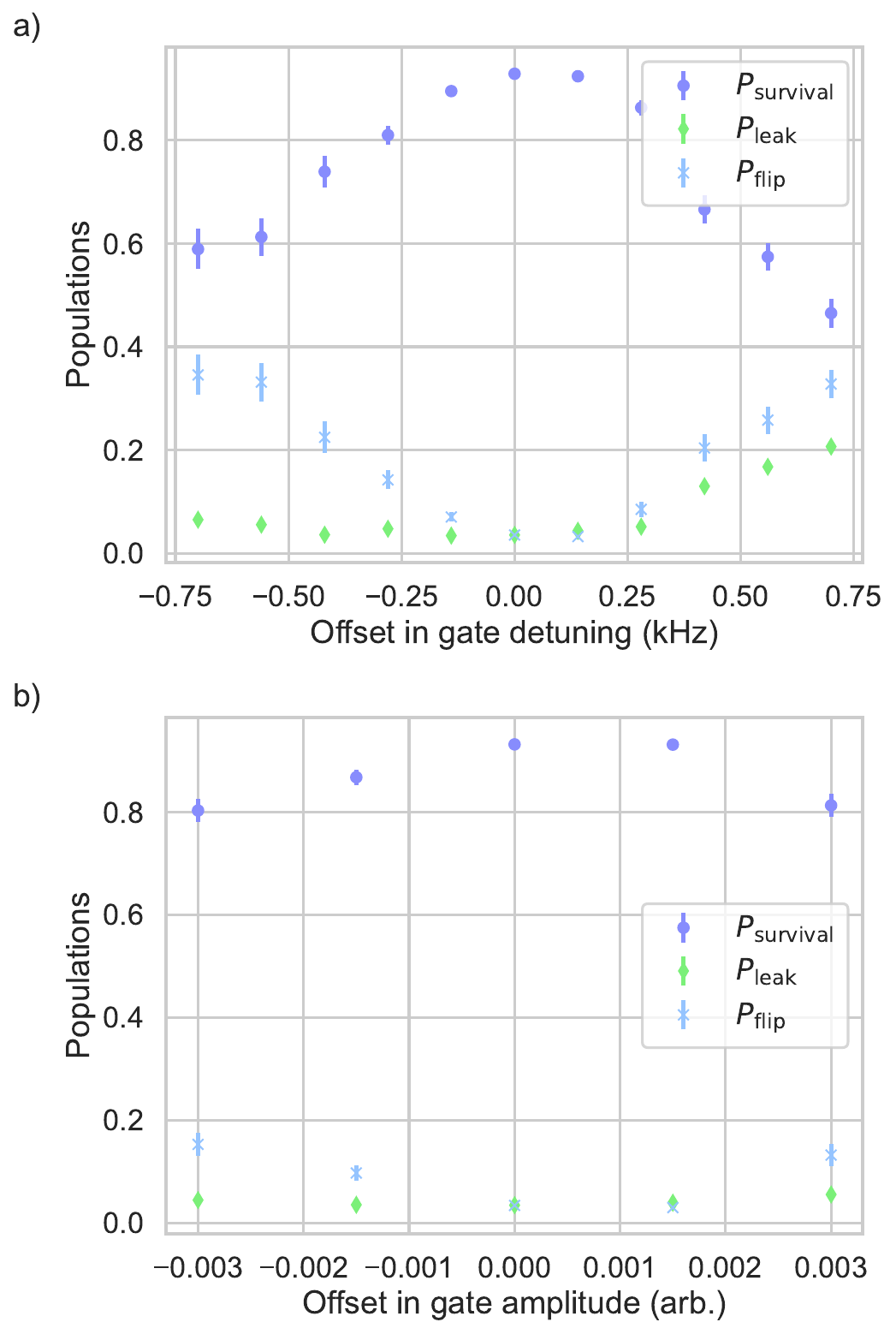}
\centering
\caption{\label{fig_calibration} Calibration data for the (a) gate detuning ($\delta$) and the (b) gate amplitude ($\Omega_\textrm{MS}$). For each of these datasets, we plot the populations versus the variation of the parameter for a fixed sequence length of 100 Cliffords with 50 randomizations and 50 shots each. Error bars indicate the 68\% confidence interval for the average populations across the different randomizations. (a) As we vary the detuning of the MS interaction from the motional mode, we pick the detuning with the lowest value of $P_\textrm{leak}$ to minimize any residual spin-motion entanglement. (b) As we vary the amplitude of the interaction, the leakage population $P_\textrm{leak}$ is roughly constant; we set the amplitude to maximize $P_\textrm{survival}$. 
}
\end{figure}

Due to its ability to rapidly and directly estimate the magnitude and nature of errors, SLERB is an excellent tool for high-fidelity MS gate calibration. Figure ~\ref{fig_calibration} shows how SLERB can be used to find the optimal value of detuning $\delta$ and gate Rabi frequency $\Omega_\textrm{MS}$. In each experiment, we apply a SLERB sequence of length $l=100$ and record populations as before, with peaks of $P_\textrm{survival}$ approximately indicating a local fidelity maximum. Following an initial coarse calibration, the high-fidelity MS gate in Fig.~\ref{fig_example} was fine-tuned through such a series of single-parameter scans, with gradually increasing length to reduce statistical uncertainty.

These calibrations also demonstrate how SLERB enables the identification of errors and independent optimization of gate parameters. With the amplitude calibration (Fig.~\ref{fig_calibration}(b)), the main effect of miscalibrating $\Omega_{\textrm{MS}}$ is a geometric phase error. Ideally, the loop closure should be unaffected as the amplitude changes and hence should not have any effect on the leakage. Indeed, for these data, $P_\textrm{leak}$ is roughly constant, and $P_\textrm{survival}$ mainly decreases due to an increase in the $P_\textrm{flip}$. Small offsets in detuning $\delta$ are mainly expected to cause geometric phase errors (spin-motion entanglement errors are first-order suppressed by Walsh modulation, thus only appear at larger offsets). This matches the experimental results in Fig~\ref{fig_calibration}(a), where $P_\textrm{leak}$ is approximately constant over $\sim 0.5$\,kHz, and then increases as expected as $\delta$ becomes less negative. 

\subsection{Verifying asymptotes}
\begin{figure}[ht]
\includegraphics[width=1\columnwidth]{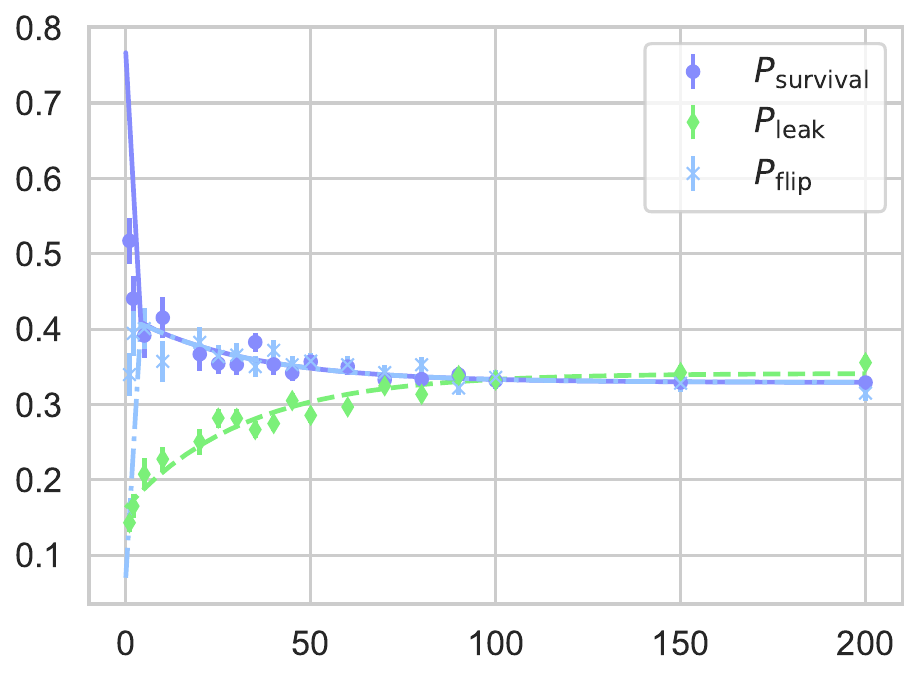}
\centering
\caption{\label{fig_asymptote} SLERB sequence with a detuning error of 2\,kHz. With this detuning offset, we observe leakage as there is residual spin-motion entanglement at the end of every gate. We fit the general decay model following Eq.~\ref{eq:blind_rb}, and find that the asymptote for $P_\textrm{leak}$ is 0.341(9). The asymptotes for $P_\textrm{survival}$ and $P_\textrm{flip}$ are 0.329(5). These values are consistent with an asymptote of 1/3 which corresponds to symmetric errors.
}
\end{figure}

As discussed in Sec.~\ref{sec:symmetry}, the error symmetry assumption can be verified by measurements of population asymptotes. In Sec.~\ref{sec:full_ms_sims}, we have confirmed numerically that, in the presence of detuning errors, the populations should asymptote to 1/3. We now confirm this experimentally by performing SLERB with $\delta$ intentionally offset by $2$\,kHz. The results are shown in Fig.~\ref{fig_asymptote}. As expected, we observe that all three populations decay to an asymptote of 1/3, confirming there no significant pumping channels into the antisymmetric state. We have also estimated through direct spectroscopy that differential qubit frequency shifts -- the main potential source of antisymmetric errors -- are smaller than 20 Hz, and thus should cause negligible gate errors ($<10^{-5}$). Thus, the symmetric error assumption is validated both theoretically and experimentally in our system.

\subsection{Applicability to laser-based MS gates}

So far, we have focused on the use of SLERB for calibrating and benchmarking laser-free MS gates. However, SLERB can be performed with laser-based gates as well, as long as two considerations are kept in mind.

First, SLERB relies on the ability to perform many sequential two-qubit gates with phase coherence of the MS phase $\phi_\textrm{MS}$. While this requirement is typically straightforward for electronic gates, the same level of coherence with laser-based approaches might be challenging. For example, if the spin phase changes with fluctuations of the optical phase, longer sequences might result in much larger errors. For two-photon stimulated Raman transitions~\cite{raman_1928, wineland_2003} at least, such fluctuations could be eliminated using phase-insensitive configurations that result in the spin phase being independent of the laser phase~\cite{lee_2005}. This configuration would result in the randomization of the motional phase of the MS interactions, which corresponds to the difference phase between the blue and red sidebands, but this phase has a negligible effect on the population decays, see Supplementary Information. Second, laser-based gates suffer from photon scattering or spontaneous emission, which can directly populate the antisymmetric subspace. Thus, additional experiments must be performed to estimate the effect of qubit decay and photon scattering on the total MS gate error (Sec.~\ref{sec:symmetry}). 

\section{Conclusion}
\label{sec:conclusion}

In conclusion, we have developed and demonstrated SLERB: a new technique for benchmarking two-qubit trapped-ion MS gates. We have validated its ability to estimate the magnitude and origin of MS gate errors through theoretical analysis and numerical simulations. Finally, we have demonstrated the use of SLERB for calibrating and characterizing high-fidelity laser-free MS gates with low statistical uncertainty. SLERB's unique selling point is that it characterizes MS gates using only MS gates. This ability to measure errors simply and directly makes it highly valuable for large-scale QCs, which will benefit from low-complexity means of measuring component-level errors. We also anticipate that SLERB can be broadly applied to other QC platforms such as superconducting qubits, as long as they are able to implement $\hat{\sigma}_\phi\hat{\sigma}_\phi$ interactions. It can also be extended to $\hat{\sigma}_z\hat{\sigma}_z$ gates~\cite{milburn_2000, leibfried_2003}, albeit with the inclusion of single-qubit rotations. The formalism developed here can also be adapted to analyze single-qubit logical randomized benchmarking over logical states $\ket{\bar{0}}$ and $\ket{\bar{1}}$ states as opposed to the physical ones considered here. The SLERB theory could be extended to more complex qubit states by incorporating leakage out of the qubit subspace altogether~\cite{chen_2025}. Finally, the use of a single two-level subspace might also be beneficial for benchmarking multi-qubit interactions for more than two qubits at a time.

\section*{Acknowledgments}
We thank members of the Quantum Performance Laboratory, especially Jordan Hines and Yale Fan, for helpful discussions and feedback on the group theory analysis. We thank Jason Saied for insight into the basis for the irreps that appears in the supplement~\cite{supplemental}. We thank Alex Kwiatkowski for insightful discussions. We thank Bryce Bjork for help setting up GAP code for the group-theoretical calculations. We thank the entire combined team at IonQ for their contributions to this work.

\bibliography{biblio}
\clearpage
\onecolumngrid
\begin{center}
\vspace{5 mm}
\textbf{\large Supplementary Information}\\
\end{center}
\twocolumngrid
\setcounter{equation}{0}
\renewcommand{\theequation}{S\arabic{equation}}

\setcounter{section}{0}


\section{Fidelity definitions}

We recall the standard definitions of average gate fidelity and process fidelity, which play a key role in this work. Below, we consider a CPTP channel $\Lambda$ and assume the state space is of dimension $d ~ (= 2^n)$ where $n$ is the number of qubits. 

The average fidelity $\bar{\mathcal{F}}(\Lambda)$ to the identity is 
\begin{equation}
    \bar{\mathcal{F}}(\Lambda) \equiv \int d \psi \bra{\psi} \Lambda(\ketbra{\psi}) \ket{\psi},
\end{equation}
where $d \psi$ represents the uniform Haar measure on the state space. One may also define the average fidelity of a quantum channel to a target unitary $U$
\begin{equation}
    \bar{\mathcal{F}}(\Lambda, U) \equiv \int d \psi \bra{\psi} U^\dagger \Lambda(\ketbra{\psi}) U \ket{\psi}. 
\end{equation}

The process (aka. entanglement) fidelity $\mathcal{F}(\Lambda)$ of a channel to the identity is formally defined
\begin{equation}
    \mathcal{F} \equiv \bra{\Psi} (\mathbbm{1} \otimes \Lambda)(\ketbra{\Psi}) \ket{\Psi}, 
\end{equation}
where $\ket{\Psi}$ is \textit{any} maximally entangled state. However, a much more useful formula can be derived~\cite{Hashim_2025}
\begin{equation}
    \mathcal{F}(\Lambda) = \frac{\Tr(\Lambda)}{d^2},
\end{equation}
where $\Lambda$ is represented as a super-operator.

The Horodecki formula~\cite{nielsen_2002} connects entanglement fidelity and average fidelity by
\begin{equation}
    \bar{\mathcal{F}}(\Lambda) = \frac{ d \mathcal{F}(\Lambda) +1 }{d + 1}.  
\end{equation}

Converting to infidelity $\bar{\mathcal{I}} = 1 - \bar{\mathcal{F}}$ and $\mathcal{I} = 1 - \mathcal{F}$, the relationship looks like 
\begin{equation}
    \bar{\mathcal{I}} = \frac{d \mathcal{I}}{d + 1}. 
\end{equation}

A convenient formula for average fidelity is~\cite{nielsen_2002}
\begin{equation}
    \bar{F}(\Lambda) = \frac{\sum_j \Tr(P_j \Lambda(P_j)) + d^2}{d^2(d+1)},
\end{equation}
where $P_j$ is a Pauli operator and the sum is over all Paulis.

Some argue~\cite{Hashim_2025} that process fidelity is more fundamental because it is ``stable'' in the sense that it composes well with growing state spaces~\cite{nielsen1996entanglementfidelityquantumerror, Hashim_2025}. If $\Lambda_1$ acts on one sub-register and $\Lambda_2$ acts on another sub-register, then the fidelity of the joint action is the product of the fidelity of the action on each sub-register:
\begin{equation}
    \mathcal{F}(\Lambda_1 \otimes \Lambda_2) = \mathcal{F}(\Lambda_1)F(\Lambda).
\end{equation}
This is not true for average gate fidelity.

\section{Generators of the benchmarking group}

The MS gate implements the following unitary:
\begin{equation*}
    \hat{U}_\text{2Q}(\theta, \phi) =\exp\Big(-\frac{i\theta_{\text{MS}}}{2}\hat{\sigma}_{\phi_{\text{MS}},1}\hat{\sigma}_{\phi_{\text{MS}},2} \Big),
\end{equation*}
where
\begin{equation*}
    \hat{\sigma}_{\phi_{\text{MS}}, j}\equiv \hat{\sigma}_{x, j}\cos(\phi_{\text{MS}})+\hat{\sigma}_{y,j}\sin(\phi_{\text{MS}}).
\end{equation*}

Now, 
\begin{align*}
    \hat{\sigma}_{\phi_{\text{MS}},1}\hat{\sigma}_{\phi_{\text{MS}},2} = \hat{\sigma}_{x, 1} \hat{\sigma}_{x, 2} \cos^2(\phi) + \hat{\sigma}_{y, 1} \hat{\sigma}_{y, 2} \sin^2(\phi) \\
    + \frac{1}{2} \sin(2 \phi) (\hat{\sigma}_{x, 1} \hat{\sigma}_{y, 2} + \hat{\sigma}_{y, 1}\hat{\sigma}_{x, 2})
\end{align*}
In matrix form, this reduces to
\begin{equation*}
    \begin{bmatrix}
        0 & 0 & 0 & e^{- 2 i \phi} \\
        0 & 0 & 1 & 0 \\
        0 & 1 & 0 & 0 \\
        e^{2 i \phi} & 0 & 0 & 0 \\
    \end{bmatrix}. 
\end{equation*}
The matrix form of the corresponding unitary is thus
\begin{equation*}
    \hat{U}_\text{2Q} = \begin{bmatrix}
        \cos\left( \tfrac{\theta}{2} \right) & 0 & 0 &-i e^{- 2 i \phi} \sin\left( \tfrac{\theta}{2} \right) \\
        0 & \cos\left( \tfrac{\theta}{2} \right) & -i \sin\left( \tfrac{\theta}{2} \right) & 0 \\
        0 & -i \sin\left( \tfrac{\theta}{2} \right) & \cos\left( \tfrac{\theta}{2} \right) & 0 \\
        -i e^{2 i \phi} \sin\left( \tfrac{\theta}{2} \right) & 0 & 0 & \cos\left( \tfrac{\theta}{2} \right) \\
    \end{bmatrix}. 
\end{equation*}

The compilations that generate the SLERB benchmarking group are
\begin{equation*}
    M_x  \equiv \hat{U}_\text{2Q}(\pi/2, 0)  = \frac{1}{\sqrt{2}} \begin{bmatrix}
        1 & 0 & 0 & -i  \\
        0 & 1 & -i  & 0 \\
        0 & -i  & 1 & 0 \\
        -i & 0 & 0 & 1 \\
    \end{bmatrix}. 
\end{equation*}
and 
\begin{equation*}
    M_y \equiv \hat{U}_\text{2Q}(\pi/2, \pi/4)  = \frac{1}{\sqrt{2}} \begin{bmatrix}
        1 & 0 & 0 & -1  \\
        0 & 1 & -i  & 0 \\
        0 & -i  & 1 & 0 \\
        1 & 0 & 0 & 1 \\
    \end{bmatrix}. 
\end{equation*}

It follows that 
\[
    M_x \cong R_x^{00, 11}(\tfrac{\pi}{2}) \oplus R_x^{01, 10}(\tfrac{\pi}{2}),
\]\[
    M_y \cong R_y^{00, 11}(\tfrac{\pi}{2}) \oplus R_x^{01, 10}(\tfrac{\pi}{2}).
\]

\section{Character table of the benchmarking group}

We used the GAP computer program to find the the character table of the benchmarking group generated by $\langle \hat{M}_x, \hat{M}_y\rangle$. A character table provides all possible irreducible representations of a group. We use it to identify the irreps that appear in our particular representation of the group benchmarking group as process matrices. The character table is given in Tab. \ref{tab:char_table}

\begin{table*}[]
    \centering
    \begin{tabular}{c|c|c|c|c|c|c|c|c|c|c|c|c|c|c|c|c}
    & 1a & 4a & 4b & 2a & 12a & 3a & 8a & 12b & 8b & 4c & 4d & 4e & 4f & 4g & 6a & 2b \\
    \hline
    X.1     & 1  & 1  & 1  & 1  & 1   & 1  & 1  & 1   & 1  & 1  & 1  & 1  & 1  & 1  & 1  & 1  \\
    X.2     & 1  & -1 & -1 & 1  & 1   & 1  & -1 & 1   & -1 & 1  & 1  & -1 & 1  & -1 & 1  & 1  \\
    X.3     & 1  & $-i$  & $i$  & -1 & -1  & 1  & $-i$  & -1  & $i$  & 1  & -1 & $-i$ & -1 & $i$  & 1  & 1  \\
    X.4     & 1  & $i$   & $-i$ & -1 & -1  & 1  & $i$   & -1  & $-i$ & 1  & -1 & $i$  & -1 & $-i$ & 1  & 1  \\
    X.5     & 2  & .  & .  & 2  & -1  & -1 & .  & -1  & .  & 2  & 2  & .  & 2  & .  & -1 & 2  \\
    X.6     & 2  & .  & .  & -2 & 1   & -1 & .  & 1   & .  & 2  & -2 & .  & -2 & .  & -1 & 2  \\
    X.7     & 2  & $1-i$   & $1+i$   & .  & $-i$  & -1 & .  & $i$   & .  & .  & $-2i$ & $-1+i$  & $2i$  & $-1-i$ & 1  & -2 \\
    X.8     & 2  & $1+i$   & $1-i$   & .  & $i$   & -1 & .  & $-i$  & .  & .  & $2i$  & $-1-i$ & $-2i$ & $-1+i$ & 1  & -2 \\
    X.9     & 2  & $-1-i$  & $-1+i$  & .  & $i$   & -1 & .  & $-i$  & .  & .  & $2i$  & $1+i$  & $-2i$ & $-1-i$ & 1  & -2 \\
    X.10    & 2  & $-1+i$  & $-1-i$  & .  & $-i$  & -1 & .  & $i$   & .  & .  & $-2i$ & $1-i$  & $2i$  & $1+i$  & 1  & -2 \\
    X.11    & 3  & -1 & -1 & -1 & .   & .  & 1  & .   & 1  & -1 & 3  & -1 & 3  & -1 & .  & 3  \\
    X.12    & 3  & 1  & 1  & -1 & .   & .  & -1 & .   & -1 & -1 & 3  & 1  & 3  & 1  & .  & 3  \\
    X.13    & 3  & $-i$  & $i$  & 1  & .   & .  & $i$   & .   & $-i$ & -1 & -3 & $-i$ & -3 & $i$  & .  & 3  \\
    X.14    & 3  & $i$   & $-i$ & 1  & .   & .  & $-i$  & .   & $i$  & -1 & -3 & $i$  & -3 & $-i$ & .  & 3  \\
    X.15    & 4  & .  & .  & .  & $-i$  & 1  & .  & $-i$  & .  & .  & $-4i$ & .  & $4i$  & .  & -1 & -4 \\
    X.16    & 4  & .  & .  & .  & $i$   & 1  & .  & $i$   & .  & .  & $4i$  & .  & $-4i$ & .  & -1 & -4 
\end{tabular}
    \caption{Character table of the benchmarking group}
    \label{tab:char_table}
\end{table*}
\hspace{-3cm}

\section{Irreduciple representations of the benchmarking group}

The benchmarking group $G$ generated by $M_x$ and $M_y$ has 96 elements. Its character table was identified using the GAP programming language~\cite{marceaux_2025_17227531}. From this character table, we identify the irreps present in a process matrix representation of the benchmarking group. We find that there are 6 irreps present in the representation, some of which appear with multiplicity. Furthermore we calculate the projectors on the irreps in the in a Pauli Transfer Matrix (PTM) basis. These irreps and their characterization are listed in Tab. \ref{tab:irrep_properties}. We further observe that there is a convenient non-canonical basis for the trivial irrep, listed in Tab. \ref{tab:trivial_irrep_decomp}.

\begin{table*}[]
    \centering
    \begin{tabular}{|c|c|c|c|c|}
        \hline
        Irrep. & Projector  & Multiplicity & $\dim(\mathcal{H}_\nu)$ & Basis \\
        \hline
        \hline
        Trivial & $\Pi_0$ & 3 & 1 & $|00\rangle \langle 00| + |11\rangle \langle 11|, |\Psi^+\rangle \langle \Psi^+|, |\Psi^-\rangle \langle \Psi^-|$, \\
        \hline
        SU(2) subspace & $\Pi_c$ & 1 & 3 & $|00\rangle \langle 11|, |11\rangle \langle 00|, |00\rangle \langle 00| - |11\rangle \langle 11|$\\
        \hline
        Symmetric-Antisymmetric transfer & $\Pi_{a, 1}$ & 1 & 1 & $|\Psi^-\rangle \langle \Psi^+|$ \\
        \hline
        Antisymmetric-Symmetric transfer  & $\Pi_{a, 2}$  & 1 & 1  & $|\Psi^+\rangle \langle \Psi^-|$\\
        \hline
        Forward mixing processes & $\Pi_{b, 1}$ & 2 & 2 & $|\Psi^+\rangle \langle 00|, |\Psi^+\rangle \langle 11|, |00\rangle \langle \Psi^-|, |11\rangle \langle \Psi^-|$ \\
        \hline
        Backwards mixing processes & $\Pi_{b, 2}$ & 2 & 2 & $|\Psi^-\rangle \langle 00|, |\Psi^-\rangle \langle 11|, |00\rangle \langle \Psi^+|, |11\rangle \langle \Psi^+|$\\
        \hline
    \end{tabular}
    \caption{Irreps. of the SLERB benchmarking group with projector labels and properties.}
    \label{tab:irrep_properties}
\end{table*}

\begin{table}[]
    \centering
    \begin{tabular}{|c|c|}
        \hline
        Irrep. & Projector\\
        \hline
        \hline
        Even-parity & $\ketbra{00}  + \ketbra{11}$ \\
        \hline
        Symmetric odd-parity & $\ketbra{\Psi^+}$ \\
        \hline
        Anti-symmetric odd-parity & $\ketbra{\Psi^-}$ \\
        \hline
    \end{tabular}
    \caption{A choice of basis for the the trivial irrep of multiplicity 3. Here $\ket{\Psi}^+ = (\ket{01} + \ket{10})/\sqrt{2}$ and $\ket{\Psi}^- = (\ket{01} - \ket{10})/\sqrt{2}$.
}
    \label{tab:trivial_irrep_decomp}
\end{table}

\section{Overlaps of irreps with computational basis states}

The overlaps between the irreps and computational basis preparation and measurements are given in Tab. \ref{tab:overlaps}.

\begin{table}[]
    \centering
    \begin{tabular}{ |cc|} 
        \hline
        $\Pi_0$ & $\begin{bmatrix}
            \tfrac{1}{2} & 0 & 0 & \tfrac{1}{2} \\
            0 & \tfrac{1}{2} & \tfrac{1}{2} & 0 \\
            0 & \tfrac{1}{2} & \tfrac{1}{2} & 0 \\
            \tfrac{1}{2} & 0 & 0 & \tfrac{1}{2} \\
        \end{bmatrix}$ \\
        \hline
        $\Pi_c$ & $\begin{bmatrix}
            \tfrac{1}{2} & 0 & 0 & -\tfrac{1}{2} \\
            0 & 0 & 0 & 0 \\
            0 & 0 & 0 & 0 \\
            -\tfrac{1}{2} & 0 & 0 & \tfrac{1}{2} \\
        \end{bmatrix}$ \\
        \hline
        $\Pi_a^1$ & $\begin{bmatrix}
            0 & 0 & 0 &0 \\
            0 & \tfrac{1}{4} & -\tfrac{1}{4} & 0 \\
            0 & -\tfrac{1}{4} & \tfrac{1}{4} & 0 \\
            0 & 0 & 0 & 0 \\
        \end{bmatrix}$ \\
        \hline
        $\Pi_a^2$ & $\begin{bmatrix}
            0 & 0 & 0 &0 \\
            0 & \tfrac{1}{4} & -\tfrac{1}{4} & 0 \\
            0 & -\tfrac{1}{4} & \tfrac{1}{4} & 0 \\
            0 & 0 & 0 & 0 \\
        \end{bmatrix}$ \\
        \hline
        $\Pi_b^1$ & $\begin{bmatrix}
            0 & 0 & 0 &0 \\
            0 & 0 & 0  & 0 \\
            0 & 0 & 0 & 0 \\
            0 & 0 & 0 & 0 \\
        \end{bmatrix}$ \\
        \hline
        $\Pi_b^2$ & $\begin{bmatrix}
            0 & 0 & 0 &0 \\
            0 & 0 & 0  & 0 \\
            0 & 0 & 0 & 0 \\
            0 & 0 & 0 & 0 \\
        \end{bmatrix}$ \\
        \hline
    \end{tabular}
    \caption{Overlaps of projectors with computational basis state preparation and measurement pairs. Columns correspond to preparations in the order 00, 01, 10, 11, and rows correspond to measurements in the same order.  }
    \label{tab:overlaps}
\end{table}

\section{Derivation of the decay model}

To derive the decays associated with the experimental implementation of SLERB in the main text, we start by analyzing a slightly more general implementation than the one performed experimentally, and then we will argue that the experimental implementation is a good approximation. Namely, we will start by assuming that the physical operations are selected randomly from the entire benchmarking group of 96 elements rather than from the subset of 24 elements that were actually chosen experimentally. After deriving decay forms in this context, we will then argue that selecting from the subset of 24 elements instead is a good approximation to the general dynamics. 

We consider the following experiment: a sequence of $l$ Cliffords are selected at random from the 96-element benchmarking group, a final $l+1$ Clifford is selected as the inverse of the product of $l$ prior gates, and the benchmarking sequence is run after preparing either $\ket{00}$ or $\ket{11}$. For concreteness, we will focus on preparing $\ket{00}$. The final state is then measured in the computational basis. To analyze the effects on outcome probabilities, we consider, as usual, gate-independent noise. That is to say, we assume that the noisy implementation $\mathcal{G}(g)$ of an element $g$ is the benchmarking group $G$ can be represented as a process matrix of the form 
\begin{equation}
    \mathcal{G}(g) = \Lambda V(g). 
\end{equation}

Now, to derive the expected decay forms, we begin with the definition of a twirl $\hat{\Lambda}_G$ of a noise channel $\Lambda$ over a group $G$: 
\begin{equation}
    \hat{\Lambda}_G= \frac{1}{|G|} \sum_{g \in G} V(g) \Lambda V(g)^{-1}. 
\end{equation}
The key insight that we exploit is that matrix powers of a twirled noise channels are equivalent to arithmetic averages over random sequences, 
\begin{equation}
    \hat{\Lambda}_G^l = \frac{1}{|G|^l} \sum_{g_l \in G}\ ... \sum_{g_1 \in G} V(g_{l+1}) \Lambda V(g_l)  \Lambda ... \Lambda V(g_1), 
\end{equation}
where $g_{l+1}$ is the inverse of the product of the $l$ prior group elements. We emphasize that this formal identification requires an approximation that the final inverse gate is perfect i.e. the error on the final Clifford gate is not twirled. In practice, the error on the final Clifford can be analyzed separately or else absorbed into measurement error. 

Next, we employ Schur's lemma (see Fact 1 in~\cite{Claes_2021}) to argue that the action of the twirled error channel is equivalent, up to a change of basis, to a block-diagonal decomposition of actions on irreps of the benchmarking group 
\begin{equation}
    \frac{1}{|G|} \sum_{g \in G} V(g) \Lambda V(g)^{-1} \cong \bigoplus_{\nu} Q_\nu \otimes \mathbbm{1}_\nu,
\end{equation}
where $\nu$ indexes the irreps of the benchmarking group, $Q_\nu$ is a $n_\nu \times n_\nu$ matrix, $n_\nu$ is the multiplicity of the irrep in the benchmarking group, and $\mathbbm{1}_\nu$ is an identity matrix of dimension $\dim(\mathcal{H}_\nu)$ equal to the dimension of the irrep. Thus,
\begin{equation}
    \hat{\Lambda}_G^l \cong \bigoplus_{\nu} Q_\nu^l \otimes \mathbbm{1}_\nu. 
\end{equation}

To analyze the dynamics of measured probabilities, we thus consider success probabilities, written in ``super-ket'' notation~\cite{Nielsen_2021}: 
\[
    \langle \langle a |  \bigoplus_{\nu} Q_\nu^l \otimes \mathbbm{1}_\nu | 00 \rangle \rangle = \sum_\nu \langle \langle a|  Q_\nu^l \otimes \mathbbm{1}_\nu | 00 \rangle \rangle,
\]
where $\langle \langle a|$ is measurement effect in the computational basis. Now, to derive the contributions to success probabilities, we make the observation that each term $Q_\nu \otimes \mathbbm{1}_\nu$ is contained in the support of the projector on the corresponding irrep. These were explicitly constructed for our benchmarking with the GAP language, see Tab. \ref{tab:irrep_properties}. We then calculate the overlap between these projectors and the relevant state preparations and measurement effects in the same basis. These results are summarized in Tab. \ref{tab:overlaps}. One may observe that the only irreps that have overlap when state preparations of 00/11 and measurement in the computational basis are the ``SU(2) subspace irrep'' and the trivial irrep. Because the SU(2) subspace irrep appears with multiplicity 1, its contribution is a single-exponential decay
\[
    \langle \langle 00 |  Q_c^l \otimes \mathbbm{1}_c | 00 \rangle \rangle = \frac{1}{2} q_{\text{RB}}^l, 
\]
where the factor of 1/2 comes from the overlap between the SU(2) irrep and this SPAM pair. Similarly, one can show
\[
    \langle \langle 11 |  Q_c^l \otimes \mathbbm{1}_c | 00 \rangle \rangle = -\frac{1}{2} q_{\text{RB}}^l. 
\]

The analysis of the trivial irrep is more subtle because it appears with multiplicity 3. The general form of the contributions follow as   
\[
    \langle \langle a |  Q_0^l \otimes \mathbbm{1}_0 | 00 \rangle \rangle = c_1 q_{\text{leak}, 1}^l + c_2 q_{\text{leak}, +}^l + c_3 q_{\text{leak}, -}^l,
\]
where $c_j$ are constants that depend on the overlap of the eigen-vectors of $Q_0$ and the SPAM pair in question and $q_j$ correspond to eigen-values. In the following section, $q_{\text{leak}, \pm}^l$ correspond to leakage rates to the two subspaces $\mathcal{S}_{\text{leak}}^{\pm}$. This form can be further simplified by invoking trace preservation, which manifests as a restriction of one eigenvalue of $Q_0$ to be 1. Hence:
\begin{equation}\label{eq:Q0}
    \langle \langle a |  Q_0^l \otimes \mathbbm{1}_0 | 00 \rangle \rangle = c_1 + c_2 q_{\text{leak}, +}^l + c_3 q_{\text{leak}, -}^l.
\end{equation}

So that the general decay forms can be derived 
\begin{align}\label{eq:gen_decay}
    \langle \langle 00 | \hat{\Lambda}_G^l | 00\rangle \rangle &= c_1 + c_2 q_{\text{leak}, +}^l + c_3 q_{\text{leak}, -}^l + \frac{1}{2} q_{\text{RB}}^l, \nonumber \\
    \langle \langle 11 | \hat{\Lambda}_G^l | 00\rangle \rangle &= d_1 + d_2 q_{\text{leak}, +}^l + d_3 q_{\text{leak}, -}^l - \frac{1}{2} q_{\text{RB}}^l.
\end{align}
We emphasize that this decay model assumes no state preparation and measurement error and no error on the final inversion gate. Without further assumptions on the nature of the noise, this general form cannot be specified further. In the next section, we will show how an assumption of leakage only within the symmetric subspace and equal leakage/seepage results in the specific decay model we use in the paper. 

The final step is to argue that the decay model above also accurately describes the dynamics of the experimental protocol in the main text -- namely where a subset of 24 elements $\mathcal{T} \subset G_{RB}$ whose action on $\mathcal{S}_{RB}$ are selected to implement the 1-qubit Clifford group on $\mathcal{S}_{RB}$. There are two ways this can be argued: 1) via a notion of asymptotic twirling or 2) through simulations. To 1) we argue that selecting from the subset of Clifford $\mathcal{T}$ provides sufficient randomness that it asymptotically generates the same dynamics as would appear if one had sampled from the group of 96 elements. This will generally only be true after some initial sequence length is passed. This is a similar approach to that of direct RB~\cite{Proctor_2019}. To 2) we have observed through numeric simulations that the decay curves observed with direct unitary simulation agree to a large degree with those found by performing the direct twirl of the same noise channel. 

\section{Decay forms with symmetric leakage}

The decay model of Eq. \ref{eq:gen_decay} represents the most specific model that can be derived with generic assumptions on the noise. However, as discussed in the main text, we expect that the dominant error channels in our system cause leakage to only the symmetric state $\mathcal{S}_{\text{leak}}^+$. Furthermore, we expect that, for the experimental regime in the main text, the dominant error mechanisms are unitary errors. This implies that the leakage rate from $\mathcal{S}_{RB}$ to $\mathcal{S}_{\text{leak}}^+$ and the seepage from $\mathcal{S}_{\text{leak}}^+$ to $\mathcal{S}_{RB}$ should be equal. We hence consider the effects on the dynamics under these assumptions. As will be shown in the next section, it turns out that these assumptions lead to a decay model that can be expressed as a continuous time Markov process. These derivations closely mirror analysis in Appendix C of Ref.~\cite{baldwin_2020}, and we refer to reader also to Sec. V of Ref.~\cite{Claes_2021} for related discussion. 

Begin by selecting projectors for the isotopic components of the trivial irrep, as in Tab. \ref{tab:trivial_irrep_decomp}, and converting them to trace-normalized superkets using the normalization convention of~\cite{baldwin_2020}:
\[
    |00\rangle \langle 00| + |11\rangle \langle 11| \mapsto \frac{|00 \rangle \rangle + |11 \rangle \rangle}{\sqrt{2}} \equiv |\mathbbm{1}_{RB}\rangle \rangle ,
\]\[
    |\Psi^\pm \rangle \langle \Psi^\pm| \mapsto |\Psi^\pm \rangle \rangle. 
\]

We make the assumption that no leakage occurs from $\mathcal{S}_{RB}$ to $\mathcal{S}_{\text{leak}}^-$, ie., that 
\begin{equation}
    \langle \langle \Psi^- |\Lambda | \mathbbm{1}_{RB}\rangle \rangle= 0.
\end{equation}
Next, define the symmetric leakage rate 
\begin{equation}
    L =  \langle \langle \Psi^+ |\Lambda | \mathbbm{1}_{RB}\rangle \rangle,
\end{equation}
and the seepage rate
\begin{equation}
    S =\langle \langle \mathbbm{1}_{RB} |\Lambda |\Psi^+ \rangle \rangle.
\end{equation}

The next step is to observe that $Q_0$ can be decomposed in this basis for the trivial irrep
\begin{equation}
    (Q_0 \otimes \mathbbm{1}_0)_{\alpha \beta} = Q_{\alpha \beta}  \langle \langle P_\alpha |\Lambda | P_j \rangle \rangle,
\end{equation}
where $P_j$ are projectors on isotopic components of the trivial irrep defined above. Based on our definitions of $L$ and $S$ we find 
\begin{equation}
    Q_{\alpha \beta} = \begin{bmatrix}
        1 - \sqrt{2} L & S & 0 \\
        L & 1 - S/\sqrt{2} & 0 \\
        0 & 0 & 1
    \end{bmatrix},
\end{equation}
where the columns correspond to $\mathcal{S}_{RB}$, $\mathcal{S}_{\text{leak}}^+$, and $\mathcal{S}_{\text{leak}}^-$, resp., and similarly for the rows. See also Eq. C5 in~\cite{baldwin_2020}. It follows that there are two eigenvalues of 1 and a nontrivial eigenvalue of $\lambda_3 = 1 - L/\sqrt{2} - \sqrt{2} S$. These are associated with the eigenvalue projectors, in order:  
\begin{equation}
    \Pi_1 = \frac{1}{L + 2 S} \begin{bmatrix}
        2 S & \sqrt{2} S & 0 \\
        \sqrt{2} L & L & 0 \\
        0 & 0 & 0,
    \end{bmatrix}
\end{equation}
\begin{equation}
    \Pi_2 =\begin{bmatrix}
        0 & 0 & 0 \\
        0 & 0 & 0 \\
        0 & 0 & 1,
    \end{bmatrix}
\end{equation}
\begin{equation}
    \Pi_3 = \frac{1}{L + 2 S} \begin{bmatrix}
        L & -\sqrt{2} S & 0 \\
        -\sqrt{2} L & 2 S & 0 \\
        0 & 0 & 0.
    \end{bmatrix}
\end{equation}

Using the overlaps between $\Pi_j$ and $|0 0\rangle \rangle$ and $|11 \rangle \rangle$ one arrives at 
\begin{equation}
    \langle \langle 00 | Q_0^l \otimes \mathbbm{1}_0|00\rangle \rangle = \frac{S}{L + 2S} + \frac{1}{2} \frac{L}{L + 2S} q_{\text{leak}, +}^l
\end{equation}
where $q_{\text{leak}, +^l }$ is the non-trivial eigenvalue $\lambda_3 = 1 - L/\sqrt{2} - \sqrt{2} S$. Furthermore, 
\begin{equation}
    \langle \langle 11 | Q_0^l \otimes \mathbbm{1}_0|00\rangle \rangle = \frac{S}{L + 2S} + \frac{1}{2} \frac{L}{L + 2S} q_{\text{leak}, +}^l. 
\end{equation}

In this way, we conclude that 
\begin{align}\label{eq:gen_decay}
    \langle \langle 00 | \hat{\Lambda}_G^l | 00\rangle \rangle &= \frac{S}{L + 2S} + \frac{1}{2} \frac{L}{L + 2S} q_{\text{leak}, +}^l + \frac{1}{2} q_{\text{RB}}^l, \nonumber \\
    \langle \langle 11 | \hat{\Lambda}_G^l | 00\rangle \rangle &= \frac{S}{L + 2S} + \frac{1}{2} \frac{L}{L + 2S} q_{\text{leak}, +}^l - \frac{1}{2} q_{\text{RB}}^l.
\end{align}

Finally, under the assumption that $L = S$, one can arrive at Eqs. \ref{eq:P_flip} and \ref{eq:P_survival}: 
\begin{align}\label{eq:gen_decay}
    \langle \langle 00 | \hat{\Lambda}_G^l | 00\rangle \rangle &= \frac{1}{3} +  \frac{1}{6} q_{\text{leak}, +}^l + \frac{1}{2} q_{\text{RB}}^l, \nonumber \\
    \langle \langle 11 | \hat{\Lambda}_G^l | 00\rangle \rangle &= \frac{1}{3} + \frac{1}{6} q_{\text{leak}, +}^l - \frac{1}{2} q_{\text{RB}}^l.
\end{align}

To connect to the Markov transition model presented in the next section, we expand $q_{\text{leak}, +}$ under the assumption $L = S$ and write
\[
    \langle \langle 00 | \hat{\Lambda}_G^l | 00\rangle \rangle = \frac{1}{3} +  \frac{1}{6} (1 - \tfrac{3}{\sqrt{2}}L)^l + \frac{1}{2} q_{\text{RB}}^l.
\]
Making a change of variables $L/\sqrt{2} \mapsto \epsilon_{\text{leak}}$, one arrives at the expression 
\[
    \langle \langle 00 | \hat{\Lambda}_G^l | 00\rangle \rangle = \frac{1}{3} +  \frac{1}{6} (1 - 3 \epsilon_{\text{leak}})^l + \frac{1}{2} q_{\text{RB}}^l,
\]
which closely matches an independent derivation from the perspective of Markov transition processes.

\section{A Markov transition model}

Under the assumption that there is no leakage to the anti-symmetric state, then the only relevant subspaces are $\mathcal{S}_{RB}$ and $\mathcal{S}_L^+$, which are both twirled over (recall $\mathcal{S}_L^+$ is 1-dimensional, so application of random phases will twirl over this space). Under these assumptions, the transfer of population between subspaces is completely incoherent. We write a Markov model that captures these dynamics, which results in a decay form that is very similar to that derived from representation-theoretical approaches. 

Thus, let us write an effective Markov model for the transition of the population 
\[
    M = \begin{bmatrix}
        1 - \epsilon_\text{RB} - \epsilon_\text{leak} & \epsilon_\text{leak} & \epsilon_\text{RB} \\
       \epsilon_\text{leak} & 1 - 2 \epsilon_\text{leak} & \epsilon_\text{leak} \\
        \epsilon_\text{RB} & \epsilon_{\text{RB}} & 1 - \epsilon_\text{RB}- \epsilon_\text{leak},
    \end{bmatrix}
\]
where $\epsilon_\text{leak}$ is a leakage rate and $\epsilon_{RB}$ is a ``spin flip rate'' that corresponds to errors in $\mathcal{S}_{RB}$.

The eigenvalues of the transfer matrix are $1, 1 - 2\epsilon_{\text{RB}} - \epsilon_{\text{leak}}, 1 - 3 \epsilon_{\text{RB}}$ and the normalized eigenvector projectors are 
\[
    P_1 = \frac{1}{3} \begin{bmatrix}
        1 & 1 & 1 \\
        1 & 1 & 1 \\
        1 & 1 & 1 \\
    \end{bmatrix},
\]\[
    P_2 = \frac{1}{2}\begin{bmatrix}
        1 & 0 & -1 \\
        0 & 0 & 0 \\
        -1 & 0 & 1
    \end{bmatrix},
\]\[
    P_3 = \frac{1}{6} \begin{bmatrix}
        1 & - 2  & 1 \\
        - 2 & 4  & - 2 \\
        1  & - 2  & 1
    \end{bmatrix}.
\]
It follows that 
\[
    M^p = \frac{1}{3}P_1 + \frac{1}{2} (1 - 2\epsilon_{\text{RB}} - \epsilon_{\text{leak}})^p P_2 + \frac{1}{6}(1 - 3\epsilon_{\text{leak}})^p P_3.
\]

Hence, preparing in 00, and measuring 00 corresponds to 
\begin{equation}
    \bra{00} M^p \ket{00} = \frac{1}{3} + \frac{1}{2} (1 - 2 \epsilon_{\text{RB}} -\epsilon_{\text{leak}})^p + \frac{1}{6} (1 - 3\epsilon_{\text{leak}})^p,
\end{equation}
and similarly for preparing in $\ket{00}$ measuring $\ket{11}$ and preparing in $\ket{00}$ measuring in the leaked state. This is essentially the exact dynamics that were derived from a representation-theoretical analysis of the problem, and indicate that a Markov description of population should be accurate under the assumption of symmetric leakage and equal rates of leakage and seepage. 

In this way we have shown
\begin{align}
P_\textrm{survival} &= \frac{1}{3} + \frac{1}{2}(1-2\epsilon_\textrm{RB}-\epsilon_\textrm{leak})^{l} + \frac{1}{6}(1-3\epsilon_\textrm{leak})^{l}, \nonumber\\
P_\textrm{leak} &= \frac{1}{3} - \frac{1}{3}(1-3\epsilon_\textrm{leak})^{l}, \nonumber\\
P_\textrm{flip} &=  \frac{1}{3} - \frac{1}{2}(1-2\epsilon_\textrm{RB}-\epsilon_\textrm{leak})^{l} + \frac{1}{6}(1-3\epsilon_\textrm{leak})^{l}, \nonumber\\
\end{align}

\noindent where $P_\textrm{survival}$ is the probability of measuring the target final state ($\ket{00}$ or $\ket{11}$), $P_\textrm{leak}$ is the probability of measuring population in the $\ket{01}$ or $\ket{10}$ states, and $P_\textrm{flip}$ is the probability of measuring the non-target state within the $\ket{00}$, $\ket{11}$ subspace. Under these assumptions of symmetric leakage and equal rates of seepage and leakage, $\epsilon_{\text{RB}}$ can be interpreted as the error per Clifford within the two-state subspace, and $\epsilon_\textrm{leak}$ as a leakage error. We note that this model was derived in the absence of state preparation and measurement errors and assumes that the final inversion gate is perfect. 


\section{Quantifying RB fidelity under partial twirling}

To derive the effective fidelity function, we analyze the decomposition of gate fidelity into its action on irreps. As emphasized in the main text, the SLERB experiment only ``fully twirls'' over the subspace $\mathcal{S}_{RB}$, and the action of the benchmarking group on the entire space $\mathcal{S}_{2Q}$ does not represent a unitary 2-design. This will complicate our derivation of the fidelity function. To perform the analysis, we will consider the effects of twirling under the benchmarking group $G_{RB}$. This Section provides the general theory, and the next section specializes to the case of SLERB benchmarking.

We begin with the definition of the channel fidelity $\mathcal{F}$ of a noise channel $\Lambda$ to the identity 
\begin{equation}
    \mathcal{F}(\Lambda) = \frac{1}{d^2}\Tr[\Lambda]. 
\end{equation}
Next, we invoke the fact that the trace is invariant under twirling to write
\begin{equation}
    \frac{1}{d^2}\Tr[\Lambda] = \frac{1}{d^2}\Tr[\hat{\Lambda}_G]. 
\end{equation}
where $d$ is the dimension of the Hilbert space. Applying a general decomposition of twirled channel, one writes 
\begin{equation}
     \frac{1}{d^2}\Tr[\hat{\Lambda}_G] = \frac{1}{d^2}\Tr\left[ \bigoplus_{\nu = 1}^l Q_\nu \otimes \mathbbm{1}_\nu \right]. 
\end{equation}
It follows for properties of the trace and tensor products that 
\begin{equation}
     \frac{1}{d^2}\Tr[\hat{\Lambda}_G] = \frac{1}{d^2} \sum_{\nu = 1} \Tr\left[Q_\nu \right]  \Tr\left[\mathbbm{1}_\nu \right]. 
\end{equation}
Now, using that $\Tr\left[\mathbbm{1}_\nu \right] =\dim(\mathcal{H}_\nu)$, it follows that
\begin{equation}\label{eq:channel_fidel_decomp}
      \mathcal{F}(\Lambda) = \frac{1}{d^2} \sum_{\nu}\dim(\mathcal{H}_\nu) \sum_{\mu} q_{\nu, \mu}, 
\end{equation}
where $q_{\nu, \mu}$ is the $\mu$th eigenvalue of the matrix $Q_\nu$.

We see above that the fidelity of a noise channel can be calculated from the knowledge of $q_{\nu, \mu}$.

\section{SLERB extended subspace fidelity}

To derive Eqs. \ref{eq:group_theory_channel_fidel} and \ref{eq:group_theory_avg_fidel}, we begin with the general decomposition of channel fidelity in terms of action on irreps, Eq. \ref{eq:channel_fidel_decomp}, reproduced below: 
\[
    \mathcal{F}(\Lambda) = \frac{1}{d^2} \sum_{\nu}\dim(\mathcal{H}_\nu) \sum_{\mu} q_{\nu, \mu}.
\]
In our case, the benchmarking group's representation decomposes into 6 irreps that are listed in Tab. \ref{tab:irrep_properties}. Hence, the general fidelity function for SLERB is 
\begin{widetext}

\begin{equation}
    \mathcal{F}(\Lambda) = \frac{1 + 3 q_\text{RB} + q_{\text{leak}, +} + q_{\text{leak}, -} + q_{a, 1} + q_{a, 2} + 2(q_{b, 1} + q_{b, 2} + q_{b, 3} + q_{b, 4}) }{16}, 
\end{equation}
\end{widetext}
where $q_{\text{RB}},  q_{\text{leak}, +}$ are the decays measured in the experiment of the main text, $q_{\text{leak}, -}$ correspond to leakage to the anti-symmetric state, and the other terms correspond to unmeasured decays that will generally manifest with coherent effects (ie. oscillations). 

In order to construct a consistent estimator of fidelity, it is necessary to measure all the decays. It would be possible to do so with an approach such as character or synthetic RB~\cite{fan2024randomizedbenchmarkingsyntheticquantum}. However, in our experiment the only decays that contribute to experimental observables are $q_\text{rb}$, $q_{\text{leak}, +}$, and $q_{\text{leak}, -}$. As we have said, we cannot resolve $q_{\text{leak}, +}$ from $q_{\text{leak}, -}$ and we expect that $q_\text{leak, -} >> q_\text{leak, +}$. To construct an approximate estimator of gate fidelity, we set $q_{\text{leak}, +} = q_{\text{leak}, -}$ (which is likely overly pessimistic) and report the unmeasured decays as the average of the measured decays. This means setting  all other terms equal to $(q_{\text{RB}} + q_{\text{leak}, +})/2$.
\begin{equation}
    \mathcal{F}(\Lambda) = \frac{1 + 8 q_{\text{RB}} + 7 q_{\text{leak}, +}}{16}.
\end{equation}
One can convert the channel fidelity above to an effective average gate fidelity via the Horodecki formula~\cite{nielsen_2002} 
\[
    \bar{\mathcal{F}} = \frac{4 \mathcal{F} + 1}{5},
\]
to arrive at 
\begin{equation}
    \bar{\mathcal{F}} =  \frac{5 + 8 q_{\text{RB}} + 7 q_{\text{leak}, +}}{20}.
\end{equation}
As emphasized in the main text, the formulas above represent an ``extended subspace fidelity'' and should not be conflated with a fully rigorous metric of two-qubit gate fidelity.  These fidelity metrics are expected to accurately describe situations where the noise is not highly biased to particular decay contributions. We calculated the $Q$-matrix across the irreps for the dominant noise channels in laser-free MS gates discussed in the main text, and explicitly verified that no significant noise bias.

\section{State Preparation and Measurement errors}

Without any state preparation and measurement (SPAM) errors, the measured populations follow

\begin{align*}
P_\textrm{survival} &= \frac{1}{3} + \frac{1}{2}(1-2\epsilon_\textrm{RB}-\epsilon_\textrm{leak})^{l} + \frac{1}{6}(1-3\epsilon_\textrm{leak})^{l}, \nonumber\\
P_\textrm{leak} &= \frac{1}{3} - \frac{1}{3}(1-3\epsilon_\textrm{leak})^{l}, \nonumber\\
P_\textrm{flip} &=  \frac{1}{3} - \frac{1}{2}(1-2\epsilon_\textrm{RB}-\epsilon_\textrm{leak})^{l} + \frac{1}{6}(1-3\epsilon_\textrm{leak})^{l}. \nonumber\\
\end{align*}

We perform Pauli-frame randomization, which averages any measurement errors of the states $\ket{0}$ and $\ket{1}$. In these experiments, we initialise both ions to $\ket{00}$. Thus, the measured populations for a sequence without any two-qubit gates are

\begin{align*}
\label{eq_spam}
P_\textrm{00} &= (1-\epsilon_\textrm{SPAM,1})(1-\epsilon_\textrm{SPAM,2}), \\ \nonumber
P_\textrm{01 or 10} &= \epsilon_\textrm{SPAM,1}(1-\epsilon_\textrm{SPAM,2})+(1-\epsilon_\textrm{SPAM,1})\epsilon_\textrm{SPAM,2}, \\
P_\textrm{11} &=  \epsilon_\textrm{SPAM, 1}\epsilon_\textrm{SPAM,2}, \nonumber\\
\end{align*}

\noindent where $\epsilon_\textrm{SPAM,i}$ is the SPAM error for ion $i$ i.e. the probability of measuring $\ket{0}$ when preparing $\ket{1}$. If $\epsilon_\textrm{SPAM,i}\ll 1$, we simplify the equations above by keeping terms only first order in $\epsilon_\textrm{SPAM,i}$. Defining the average state preparation error 

\begin{align}
\bar{\epsilon}_\textrm{SPAM}=\frac{1}{2}(\epsilon_\textrm{SPAM,1}+\epsilon_\textrm{SPAM,2}),
\end{align}

\noindent and using the initial populations to calculate the decays, we obtain

\begin{align}
\label{eq:slrb_decay_spam}
P_\textrm{survival} &= \frac{1}{3} (1- \bar{\epsilon}_\textrm{SPAM}) \nonumber\\
&+\frac{1}{2}(1-2\bar{\epsilon}_\textrm{SPAM})(1-2\epsilon_\textrm{RB}-\epsilon_\textrm{leak})^{l} \nonumber \\
&+ \frac{1}{6}(1-4\bar{\epsilon}_\textrm{SPAM})(1-3\epsilon_\textrm{leak})^{l}, \nonumber\\
P_\textrm{leak} &= \frac{1}{3} (1 + 2\bar{\epsilon}_\textrm{SPAM}) - \frac{1}{3}(1-4\bar{\epsilon}_\textrm{SPAM})(1-3\epsilon_\textrm{leak})^{l}, \\
P_\textrm{flip} &=  \frac{1}{3} (1- \bar{\epsilon}_\textrm{SPAM}) \nonumber \\
&- \frac{1}{2}(1-2\bar{\epsilon}_\textrm{SPAM})(1-2\epsilon_\textrm{RB}-\epsilon_\textrm{leak})^{l} \nonumber \\ 
&+ \frac{1}{6}(1-4\bar{\epsilon}_\textrm{SPAM})(1-3\epsilon_\textrm{leak})^{l}. \nonumber
\end{align}

\noindent This derivation accounts for the starting states $\ket{01}$ and $\ket{10}$ having different asymptotes to $\ket{00}$ or $\ket{11}$. There will be an additional correction to the asymptotes due to measurement errors, despite Pauli-frame randomization, as we stay within the symmetric subspace. For simplicity, we ignore those here, as the main data is far from the asymptotes, and any correction would be a fraction of the measurement error, which is less than $10^{-4}$.

\section{Effect of motional state}

\begin{figure}[!h]
\includegraphics[width=1\columnwidth]{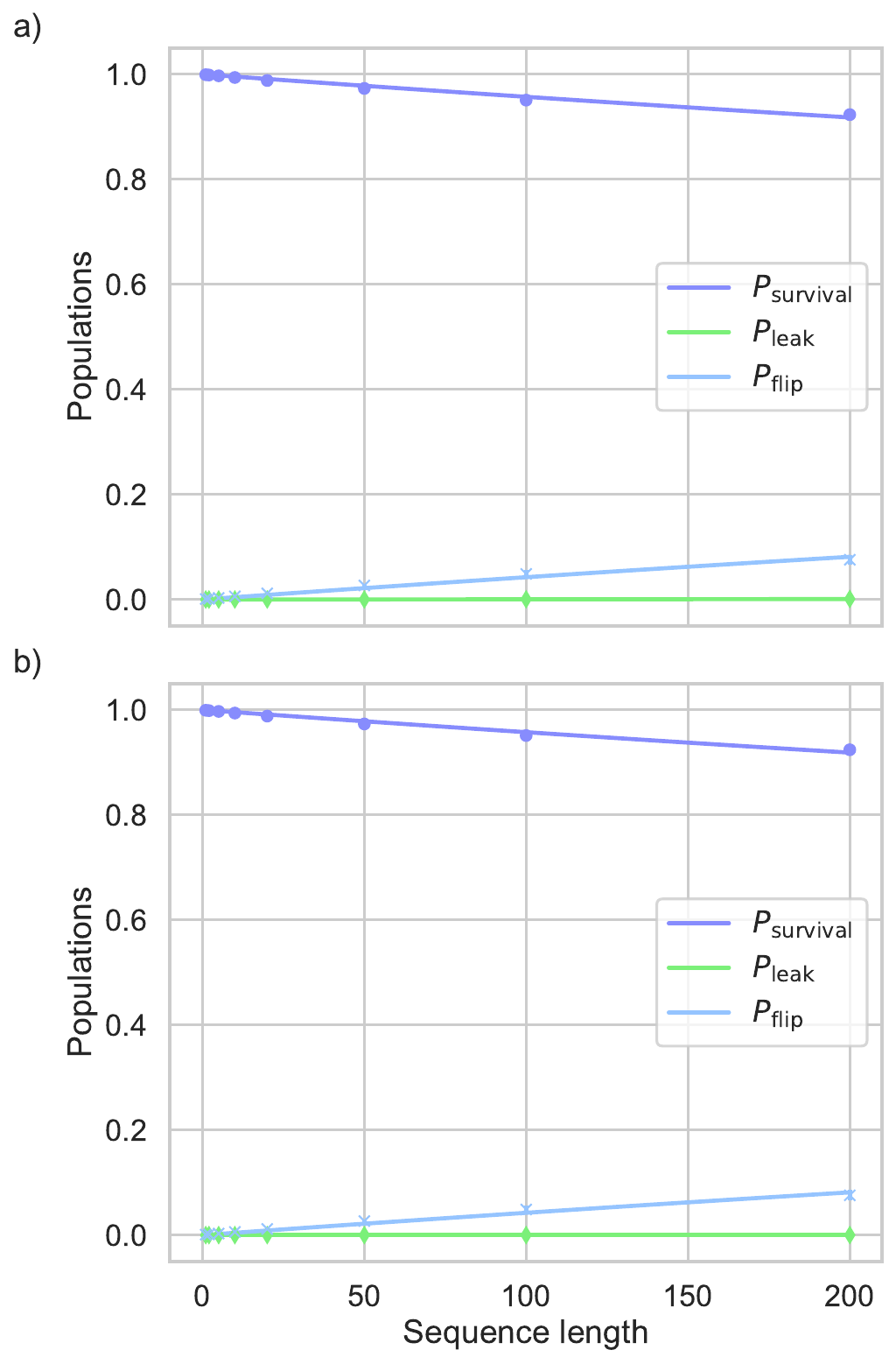}
\centering
\caption{\label{fig_ms_sim_det_small} Numerical simulations of SLERB sequences for a fractional detuning error of $10^{-2}$ with a) the motional state unchanged or b) reset to the ground state after every two-qubit gate. For each sequence length, we sample 50 random sequences; the error bars correspond to the 68\% confidence interval for the average populations from all the randomisations. The solid lines correspond to fits to the data following Eq.~\ref{eq_slerb_decay}. For this error, the motional phase has a negligible effect on the dynamics or the estimated two-qubit gate error of $2.5\times10^{-4}$.
}
\end{figure}

\begin{figure}[!h]
\includegraphics[width=1\columnwidth]{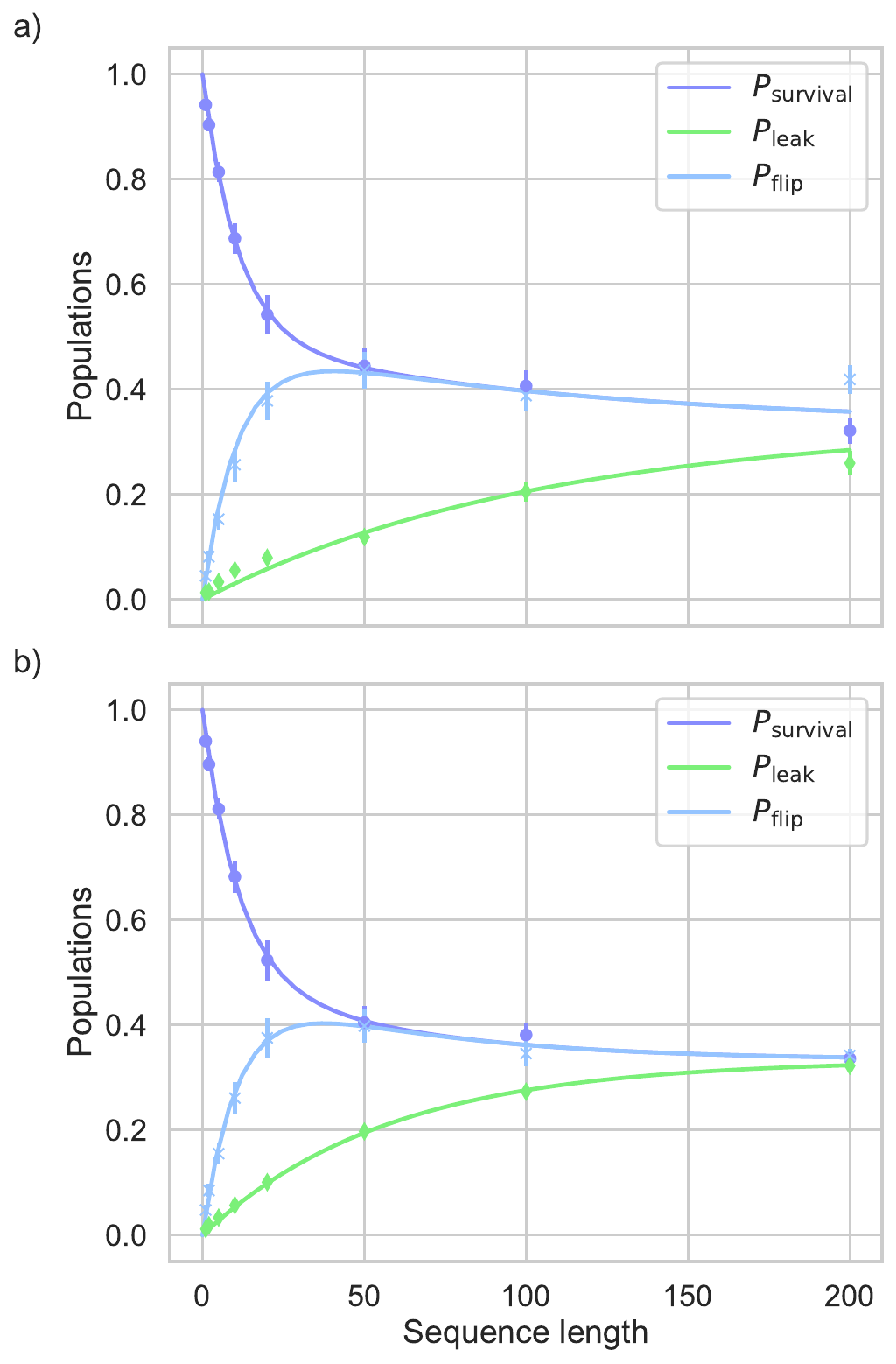}
\centering
\caption{\label{fig_ms_sim_det} Numerical simulations of SLERB sequences for a fractional detuning error of $7\times10^{-2}$ with a) the motional state unchanged or b) reset to the ground state after every two-qubit gate. For each sequence length, we sample 50 random sequences; the error bars correspond to the 68\% confidence interval for the average populations from all the randomisations. The solid lines correspond to fits to the data following Eq.~\ref{eq_slerb_decay}. In contrast to Fig.~\ref{fig_ms_sim_det_small}, the motional state has a more significant effect on the dynamics and deviates slightly from the model. However, both sets of data result in an estimated two-qubit gate error of $2.5\times10^{-2}$.
}
\end{figure}

Errors in the MS interaction can result in residual spin-motion entanglement (SME) at the end of the gate duration. Without additional ground-state cooling, any subsequent two-qubit gates will be performed with this residual SME, which can result in non-Markovian errors due to a coherent build up of motional population. For the theory and experiments described in the main text, we do not consider this effect. Here, we perform simulations, again in QuTiP, to investigate this effect in two regimes: a fractional detuning error of $10^{-2}$ (Fig.~\ref{fig_ms_sim_det_small}), and $7\times10^{-2}$ (Fig.~\ref{fig_ms_sim_det}). The first case corresponds to a two-qubit gate error of $2.5 \times 10^{-4}$, i.e. illustrates a scenario where the vast majority of the experimentally measured gate error is due to motional errors. For each of these cases, we perform simulations where we either keep the motional state unchanged after every gate sequence, or reset it to the ground state after every two-qubit gate. For the smaller error, the motional state has a negligible effect on the fidelity estimate. Thus, we conclude that our experimental fidelity estimation is not significantly biased by coherent motional population buildup. For the larger error, this effect is still small, but we observe the data deviating from the model as shown in Fig.~\ref{fig_ms_sim_det}a.

\label{suppl_material}

\end{document}